\documentclass[%
 reprint,
superscriptaddress,
 amsmath,amssymb,
 aps,
]{revtex4-2}

\usepackage{graphicx}% Include figure files
\usepackage{dcolumn}% Align table columns on decimal point
\usepackage{bm}% bold math
\usepackage{color}
\newcommand{\addihpc}{Institute of High Performance Computing, A$^\star$STAR (Agency for Science, Technology and Research), Singapore}
\newcommand{\addeinhoven}{Department of Applied Physics and Science Education, Eindhoven University of Technology, 5600MB, The Netherlands}

\newcommand{\addsutd}{Science, Mathematics and Technology Cluster, Singapore University
of Technology and Design, Singapore}

\newcommand{\addanu}{Nonlinear Physics Center, Research School of Physics, Australian National University, ACT 2601 Australia}

\begin{document}

\preprint{APS/123-QED}

\title{Mie-tronics supermodes and symmetry breaking in nonlocal metasurfaces}% Force line breaks with \\

\author{Thanh Xuan Hoang}
\email{hoangxuan11@gmail.com}
\affiliation{\addihpc}%
%\affiliation{\addindependent}
\thanks{Current address: Independent Researcher, Oakville, Ontario, Canada}

\author{Ayan Nussupbekov}
% \email{hoangtx@ihpc.a-star.edu.sg}
\affiliation{\addihpc}%

\author{Jie Ji}
\affiliation{\addeinhoven}%
\author{Daniel Leykam}%
%\email{daniel.leykam@gmail.com}
\affiliation{\addsutd}%

\author{Jaime G\'omez Rivas}
\affiliation{\addeinhoven}

\author{Yuri Kivshar}%
\email{yuri.kivshar@anu.edu.au}
\affiliation{\addanu}%

\date{\today}

\begin{abstract}
It is usually believed that symmetry breaking in photonic systems leads to weaker confinement, such as in the case of metasurfaces when bound states in the continuum are replaced by quasi-bound states with lower quality factors ($Q-$factors). Here, in contrast,  we reveal that symmetry breaking can instead enhance light trapping by strengthening in-plane nonlocal coupling pathways. We consider finite-size arrays of optical resonators supporting Mie resonances (a Mie-tronics platform) and employ diffraction and multiple-scattering analyses to demonstrate that diffractive bands and Mie-tronics supermodes originate from the same underlying Mie resonances but differ fundamentally in their physical nature. Finite arrays exhibit the $Q$-factor enhancement driven by redistributed radiation channels, thus reversing the trends predicted for infinite lattices. We demonstrate that controlled symmetry breaking opens new electromagnetic coupling channels, enabling polarization conversion in nonlocal metasurfaces. These novel findings establish a unified wave-physics platform linking both scattering and diffraction theories. Also, they outline the design principles for multifunctional metasurfaces that exploit nonlocality for advanced light manipulation, computation, and emission.
\end{abstract}

\maketitle

\section{Introduction}
The field of optical metasurfaces is rapidly evolving from fundamental science to practical technologies, enabling compact platforms for imaging, sensing, energy harvesting, and information processing~\cite{Brongersma2025second}. Conventional metasurfaces comprise weakly interacting subwavelength elements, allowing a local description. Yet, recent analyses of their limits show that surpassing these constraints demands nonlocal design~\cite{Cai2020inverse,Shastri2023nonlocal,Yao2024nonlocal}. When inter-element coupling becomes strong, collective optical excitations emerge. These excitations result in nonlocal metasurfaces governed by coherent modes extending across the structure. Because these modes are typically excited by plane waves, they can often be interpreted through diffraction theory, earning the name diffractive nonlocal metasurfaces~\cite{Overvig2022diffractive}. Understanding such systems benefits from revisiting classical diffraction theory.

In his 1907 ``Dynamical Theory of Gratings'', Rayleigh extended Fresnel’s scalar treatment of reflection and diffraction to periodic metallic surfaces. He noted that Fresnel’s local approximation fails when the grating period approaches the wavelength. In this regime, the recesses of the grating ``act as resonators”~\cite{Rayleigh1907dynamical}. Rayleigh thus recognized, decades ahead of modern electromagnetic modeling, that resonance fundamentally alters diffraction and that a “more strictly dynamical theory” is required. Ugo Fano later provided such a framework~\cite{Fano1941theory}. He introduced coupling among diffracted and evanescent orders, showing how these “Rayleigh resonators” form quasi-stationary surface waves. The interference between these surface waves and the continuum gives rise to asymmetric Fano lineshapes. The Rayleigh–Fano formalism laid the foundation for modern numerical solvers such as the rigorous coupled-wave analysis (RCWA), also known as the Fourier modal method~\cite{Liu2012s4}. These tools are used here to compute diffractive bands of nonlocal metasurfaces.

The design of nonlocal metasurfaces often begins with idealized infinite arrays of resonators, where symmetry breaking within the unit cell serves as a versatile tool for tailoring the dispersion and radiation properties of diffractive bands~\cite{Nguyen2018symmetry,Overvig2020selection,You2024resonance,Barreda2025bound}. Engineering such diffractive bands has recently fueled advances in analog optical computing and wave-based information processing~\cite{Guo2018photonic,Kwon2018nonlocal,Komar2021edge,Liu2024edge}. In practical implementations, however, metasurfaces are finite—a feature that has drawn growing attention~\cite{Ho2024finite,Karavaev2025emergence,Chen2025observation,Zhang2025vortex,Jie2025near}. While compactness makes them appealing for integration, it simultaneously invalidates the assumption of infinite periodicity inherent to diffraction theory. As the array size decreases, translational symmetry is broken and the momentum conservation underlying Fourier modal methods no longer holds. In finite arrays, collective resonances must be treated explicitly. Such resonances are discrete, inherently nonlocal, and strongly governed by the global geometry of the structure. In this regime, multiple scattering provides a more appropriate physical description~\cite{Lagendijk1996resonant}.

At the microscopic level, these collective effects originate from interactions among Mie modes—localized resonances of individual dielectric or metallic resonators that form the first key element of the Mie-tronics framework~\cite{Rybin2024metaphotonics}. The coupling coefficients of these modes constitute the second key element of Mie-tronics~\cite{Chew1993efficient}. Within a multipole-expansion formalism, these coefficients quantify how the resonances interact across the metasurface. Mie-tronics therefore provides a unified framework describing both the local response of isolated resonators~\cite{Li2024machine,Barati2025toward} and the nonlocal response of arrays of coupled resonators. These collective supermodes can enhance device performance and enable new design concepts~\cite{Krasnok2012all,Babicheva2024mie,Ha2024optoelectronic,Mao2025lateral}.

Here, we systematically study how symmetry breaking affects both diffractive and Mie-tronics supermodes. Contrary to expectations from diffraction bands of infinite arrays, symmetry breaking in finite metasurfaces can enhance in-plane multiple scattering and increase the quality factor ($Q-$factor) of certain nonlocal supermodes. We further demonstrate that coupling among Mie modes produces additional bonding and anti-bonding supermodes beyond the Fourier-mode couplings of classical diffraction theory. This effect expands the functional landscape of nonlocal metasurfaces. Notably, anti-bonding supermodes remain robust in the presence of a quartz substrate, whereas bonding supermodes are suppressed. Finally, we show that distinct unit-cell geometries support identical Mie-resonant supermodes. This result highlights the robustness and design versatility of the Mie-tronics framework for both diffractive and finite-size nonlocal metasurfaces.

%%%%%%%%%%%%%%%%%%%%%%%%%%%%%

\section{Fundamentals of Mie-tronics} 
Mie-tronics has evolved from early developments in 19th-century scattering theory to a modern framework for describing light–matter interactions in complex nanophotonic structures. A prominent example is resonant metasurfaces, where arrays of particles support collective optical modes. This section introduces the key concepts required to describe light interaction with individual particles as well as particle arrays. We first review the historical origins of multipole scattering theory, then clarify the distinction between diffraction-based and scattering-based descriptions. Finally, we introduce Mie modes and multipole coupling as the building blocks of a design framework for advanced photonic systems.

\subsection{Historical origins of Mie-tronics}
The intellectual roots of Mie-tronics trace back to the 19th-century work of Clebsch. At a time when lens and mirror design was largely governed by geometrical principles, he sought a rigorous wave-based framework for optical devices~\cite{Logan1965survey}. Using both plane-wave and point-like sources, he introduced an early form of the multipole expansion to solve boundary conditions—an approach that later became a cornerstone of scattering theory. Although Clebsch ultimately declared his effort a failure, his pioneering use of the multipole expansion has become a powerful tool for describing wave interactions with structured media~\cite{Novotny2012principles}. What once seemed unattainable is now well established: the laws of reflection and refraction follow directly from wave physics~\cite{Bohren2008absorption}. In particular, these laws can be derived from the scattering coefficients of spherical boundaries~\cite{Hoang2012interpretation,Hoang2013rigorous}. In this sense, Clebsch’s attempt to reinterpret bulky optical elements in wave terms resonates with today’s drive to realize ultrathin metasurfaces that harness wave phenomena to replace conventional optics.

Clebsch’s ideas influenced Lorenz, who first derived the exact solution for plane-wave scattering by a sphere. This result, however, later became more widely associated with Mie because his systematic multipole analysis provided the analytical foundation for modern scattering theory~\cite{Wriedt2012mie}. Unlike Lorenz, Mie was primarily inspired by Rayleigh, who had independently investigated acoustic and optical scattering problems similar to those considered by Clebsch. Notably, Mie introduced the now-standard term ``Rayleigh scattering” to distinguish the point-particle regime from his treatment of larger spheres~\cite{Mie1976contributions}.

Over the following century, these contributions coalesced into the framework of generalized Lorenz–Mie theory, emphasizing both the intrinsic Mie modes of scatterers and the multipole content of the excitation sources~\cite{Gouesbet2011generalized}. Advances in nanoscience have extended scattering analysis to particles of arbitrary shape~\cite{Kildishev2025art}, enabling new classes of photonic devices~\cite{Dorodnyy2023mie}. This convergence of multipole physics and nanophotonics defines the emerging paradigm of Mie-tronics~\cite{Won2019into,kivshar_Mie}—a framework that unifies classical scattering theory with modern metasurface design.

\subsection{Diffraction and scattering}

The early development of scattering theory was closely intertwined with the study of diffraction. In the 19th century, Clebsch, Lorenz, and Rayleigh often described light–particle interactions in the language of diffraction~\cite{strutt1871lviii}.  
Before Maxwell’s electromagnetic formulation, when the Huygens–Fresnel principle dominated, the terms ``diffraction'' and ``scattering'' were used almost interchangeably.  
With the establishment of Maxwell’s equations, however, diffraction came to be recognized as a specific manifestation of the broader phenomenon of electromagnetic scattering~\cite{Rayleigh1881x,Feynman2015feynman}.

Even today, these terms persist in distinct theoretical contexts, sometimes leading to ambiguity. For example, in diffractive nonlocal metasurfaces, ``scattering'' is often used to describe the radiative leakage of bound states that fundamentally originate from diffraction processes~\cite{Hoang2025collective}. For metasurfaces with subwavelength period, classical diffraction theory typically considers coupling through a single Fourier order as the only radiation channel. However, electromagnetic waves support multiple scattering pathways because of their positive-definite energy spectrum~\cite{Lagendijk1996resonant}. This fundamental property makes trapping light more difficult than trapping electrons, which is readily achieved in negative potentials, such as quantum wells~\cite{John1991localization}. In Mie-scattering systems, strong confinement therefore requires high-index particles that support multipolar resonances. These additional channels, associated with multipolar displacement currents induced in the scatterers, extend beyond the standard Fourier harmonics~\cite{Feng2025beyond}. Accounting for them is useful for engineering superlattices with tailored spectral responses~\cite{Song2025emergence}. In this multipolar scattering picture, high-$Q$ nonlocal supermodes arise from constructive interference among several multipole radiation channels rather than from suppressing scattering itself.

For conceptual clarity, we adopt the following convention. Results obtained using RCWA are referred to as diffraction theory. In this approach, electromagnetic fields are expanded into Fourier harmonics under plane-wave excitation and in-plane periodic boundary conditions. This formulation originates from the classical grating analyses of Rayleigh and Fano~\cite{Rayleigh1907dynamical,Fano1941theory}. In contrast, results derived from multipole expansions of fields scattered by the metasurface under dipole excitation are referred to as scattering theory. These calculations are performed under open (radiative) boundary conditions. This framework is rooted in the pioneering works of Clebsch, Lorenz, Rayleigh, and Mie~\cite{Logan1965survey}.

As shown below, the Mie-mode basis provides a clear connection between diffraction and scattering theories. In this unified picture, Mie-tronics bridges Fourier-mode analyses with multipole expansions. It reveals their common foundations and provides a systematic approach for designing nonlocal multipole-based metasurfaces.

\subsection{Mie modes as the building blocks}

\begin{figure}[hbtp]
\includegraphics[width = 8 cm]{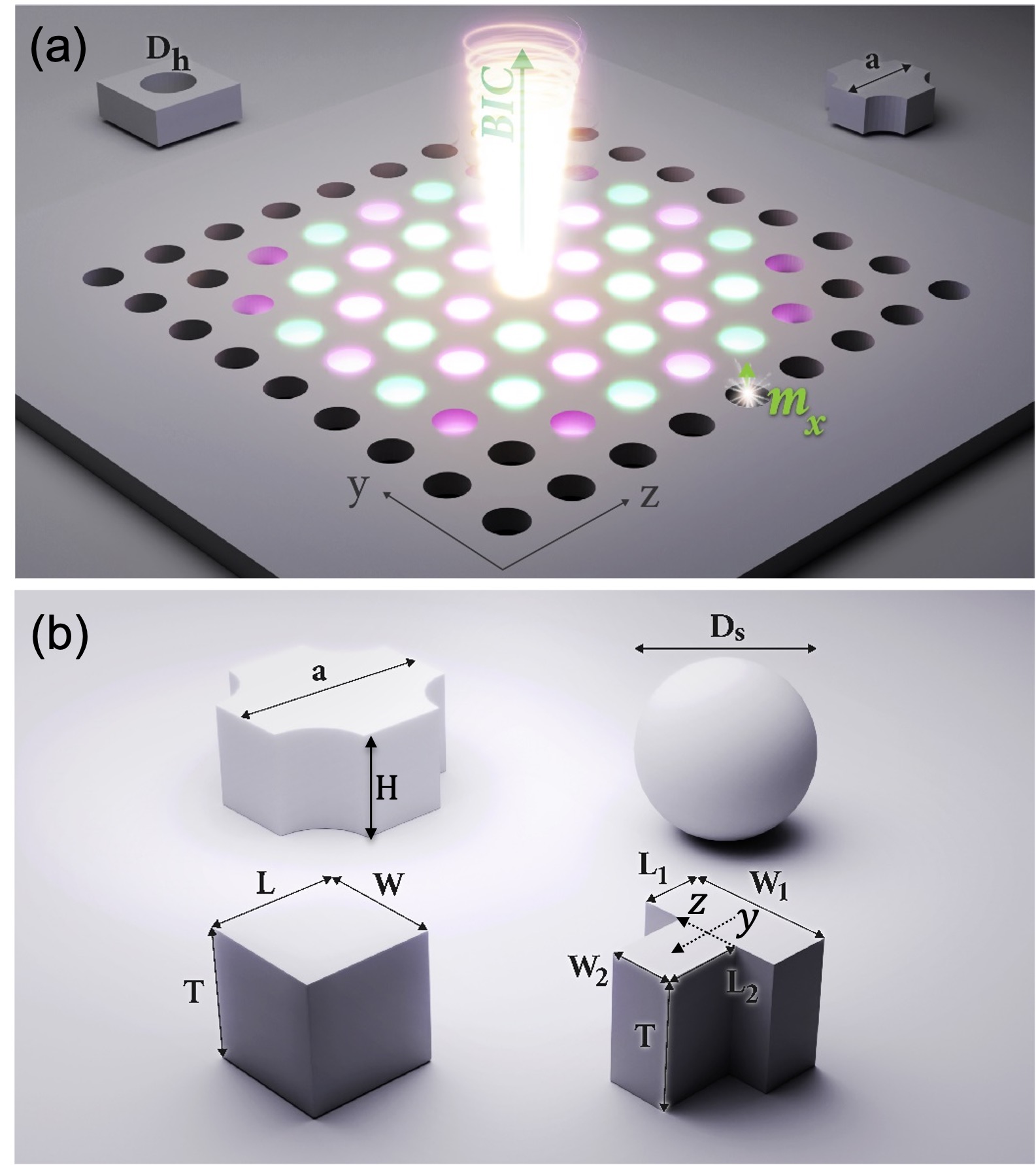}% Here is how to import EPS art
\caption{\label{F1} Beyond spheres: equivalent unit cells in the Mie-tronics framework.
(a) Schematic of a magnetic dipole ($m_x$) interacting with an array of air holes in a silicon slab. The insets show two unit cells that are equivalent from a photonic-crystal perspective but distinct in the Mie-tronics description.
(b) Four functionally equivalent unit cells with hole, sphere, square, and T-shaped geometries. Symmetry breaking of the T-shaped unit cell along the $z$ axis redistributes radiation channels, enabling additional coupling pathways.}
\end{figure}

Although practical scattering systems are more complex than a single sphere, Mie coefficients and multipole expansions remain fundamental tools for understanding their optical response. We therefore briefly review the standard Mie formalism, its implementation, and physical meaning. In this framework, the incident field on a sphere ${\bf E}^{\text{inc}}$, scattered field by the sphere ${\bf E}^{\text{sct}}$, and internal field in the sphere ${\bf E}^{\text{int}}$ are expanded in a multipole basis:
\begin{align*}
{\bf E}^{\text{inc}}({\bf r})
&=\sum_{l=1}^{L_{\text inc}}\sum_{m=-l}^{l}\left[p_{lm}{\bf N}_{lm}(k{\bf r})+q_{lm}{\bf M}_{lm}(k{\bf r})\right],\\
{\bf E}^{\text{sct}}({\bf r})
&=\sum_{l=1}^{L_{\text R}}\sum_{m=-l}^{l}\left[a_lp_{lm}{\bf N}^{(1)}_{lm}(k{\bf r})+b_lq_{lm}{\bf M}^{(1)}_{lm}(k{\bf r})\right],\\ 
{\bf E}^{\text{int}}({\bf r})
&=\sum_{l=1}^{L_{\text R}}\sum_{m=-l}^{l}\left[c_lp_{lm}{\bf N}_{lm}(n_{\text s}k{\bf r})+d_lq_{lm}{\bf M}_{lm}(n_{\text s}k{\bf r})\right],
\end{align*}
where $a_l$ and $b_l$ ($c_l$ and $d_l$) are the external (internal) Mie coefficients, determining how each multipole component radiates into free space or induces displacement currents within the sphere. Here $n_{\text s}$ denotes the refractive index of the sphere and $k=2\pi/\lambda$ the free-space wavenumber. 

When the incident field has an analytical multipole representation with order $L_{\text{inc}}$, the solution is fully analytical. In the original Mie formulation, however, the series extends to infinity and must be truncated at a finite order $L_{R}$, yielding a semi-analytical solution. An empirical rule for a sphere of radius $R$ is~\cite{Hoang2017fano}:
\begin{equation}
 L_{\text R}\approx 2\pi R/\lambda+4.05\sqrt[3]{2\pi R/\lambda}+2.\label{E1}
\end{equation}
For example, the sphere with parameters presented in Table~\ref{T1} requires $L_{\text{R}}$ between 7 and 10 for wavelengths near 1500 nm. While our focus is on magnetic dipole (MD, $l=1$) interactions, setting $L_{\text{R}}=1$ is inadequate because it neglects higher-order coupling, leading to inaccurate estimates of $Q-$factors and field enhancements. Moreover, when $L_{\text inc}>L_{\text R}$, the higher-order incident components ($l>L_{\text R}$) cannot couple to internal Mie modes and thus propagate unscattered.

Mie coefficients are formally valid only for isotropic, homogeneous spheres. For arbitrary geometries—such as the hole, square, and T-shaped unit cells in Fig.~\ref{F1}—full-wave simulations are required. We therefore employ the commercial software Lumerical FDTD to evaluate the $Q-$ and Purcell factors. The Purcell factor is defined as $P/P_0$, where $P$ and $P_0$ denote the power dissipated by a dipole source in the presence of the metasurface and in the corresponding homogeneous medium, respectively. The $Q-$factor is defined as $\lambda/\Delta\lambda$, where $\lambda$ is the resonance wavelength and $\Delta\lambda$ is the full width at half-maximum of the spectral response.
The reflectance, computed using Lumerical RCWA, is defined as the ratio of reflected to incident power integrated over a unit cell. Further details on their formal definitions and derivations are provided in our previous studies~\cite{Hoang2017fano,Hoang2025collective}. All FDTD project files, simulation setups, and MATLAB scripts used in this work are publicly available on Zenodo~\cite{Zenodo}.

\begin{table*}[hbtp]
\caption{\label{T1} Geometrical parameters of the unit cells shown in Fig.~\ref{F1}.}
\begin{ruledtabular}
\begin{tabular}{cccccc}
Unit cell & Period (nm) & Diameter / width (nm) & Thickness (nm) & Material & Environment \\
\hline
Sphere   & 425 & $D_s = 420$ & --- & Si ($n=3.5$) & Vacuum \\
Hole    & 529 & $D_h = 350$ & $H = 600$ & Si slab ($n=3.5$) & Vacuum \\
Square   & 720 & $W = L = 400$ & $T = 500$ & Si ($n=3.5$) & Vacuum \\
T-shape  & 720 & $W_1,\,L_1,\,W_2,\,L_2 = 400,\,240,\,280,\,160$ & $T = 500$ & Si ($n=3.5$) & Vacuum \\
\end{tabular}
\end{ruledtabular}
\end{table*}

Figure~\ref{F1}(a) shows a schematic of a photonic-crystal slab interacting with an $x$-oriented MD emitter. This system can produce a vortex beam with suppressed radiation in the normal direction, a key feature discussed in this work. The suppression originates from destructive interference between displacement multipolar currents induced in the system. These multipolar currents collectively form an eigenmode whose fields are symmetry-mismatched with the normal radiation channel. Within diffraction theory, this eigenmode is commonly identified as a symmetry-protected bound state in the continuum (BIC)~\cite{Hoang2025collective}. In Mie-tronics, this BIC manifests as a vortex beam~\cite{Chen2022observation}.

From a Mie-tronics perspective, the unit cell shown in the right inset of Fig.~\ref{F1}(a) is functionally equivalent to the three other unit cells in Fig.~\ref{F1}(b). Arrays of these geometries are adapted from previous studies~\cite{Chen2022observation,Liu2019high,Hoang2024photonic}, with their parameters summarized in Table~\ref{T1}. Despite differences in size and shape, we show that these arrays support the same collective MD resonances. Geometry transformations and symmetry breaking of the unit cells, however, modify the electric and magnetic field distributions of the Mie modes and open additional coupling pathways for controlling light–matter interactions.

Before addressing other geometries, we first examine the spectral profiles of the Mie coefficients and how supermodes (collective resonances) arise in finite arrays of spheres, as well as their connection to Bloch-band formation.

\begin{figure}[hbtp]
\includegraphics[width = 6 cm]{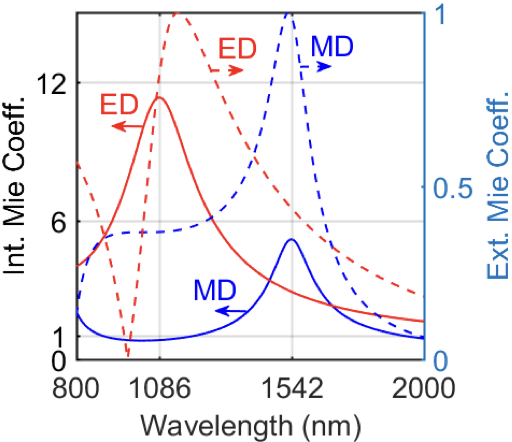}
\caption{\label{F2} 
Spectral dependence of the lowest-order Mie coefficients of the silicon sphere listed in Table~\ref{T1}. Internal (left) and external (right) electric-dipole (ED) and magnetic-dipole (MD) coefficients reveal broad overlapping resonances.}
\end{figure}
Figure~\ref{F2} shows the spectral profiles of the two lowest-order Mie coefficients: the electric (ED) and magnetic dipoles. The internal ED and MD resonances exhibit broad spectral distributions. For a sphere with unit index contrast, the internal coefficients equal 1; thus, values exceeding unity indicate enhanced energy storage. The ED mode spans nearly the entire wavelength range shown, while the MD mode extends from 1200 to 1940 nm. 

This broad spectral overlap explains why single-mode models, such as the dipole approximation, cannot accurately capture the $Q-$ and Purcell factors of optically coupled arrays. In these systems, multiple multipoles contribute to the scattering process that governs wave behavior~\cite{Hoang2024photonic}. Their contributions are set partially by their intrinsic spectral profiles. The role of spectral overlap in Mie-tronics is discussed further in Subsection~\ref{Sec2E}.

Despite their broad spectra, the ED and MD resonances correspond to whispering-gallery modes, consistent with both geometrical and wave-optics descriptions. Early photonic-crystal studies argued that large unit cells would not support photonic bands due to the dominance of geometrical optics~\cite{John1991localization}. Mie-tronics revises this view: all Mie modes inherently embody wave physics and can couple collectively. As shown in prior work, coupling between low- and high-order modes gives rise to photonic bands~\cite{Hoang2017fano,Hoang2020collective}. 

Next, we show that coupling these intrinsic Mie modes results in delocalized collective modes—Mie-tronics supermodes—in finite arrays. We then examine how these supermodes connect to Bloch bands in infinite periodic systems.

\subsection{Beyond Bloch bands: Anti-bonding and bonding supermodes}
\begin{figure*}[htbp]
\includegraphics[width = 16 cm]{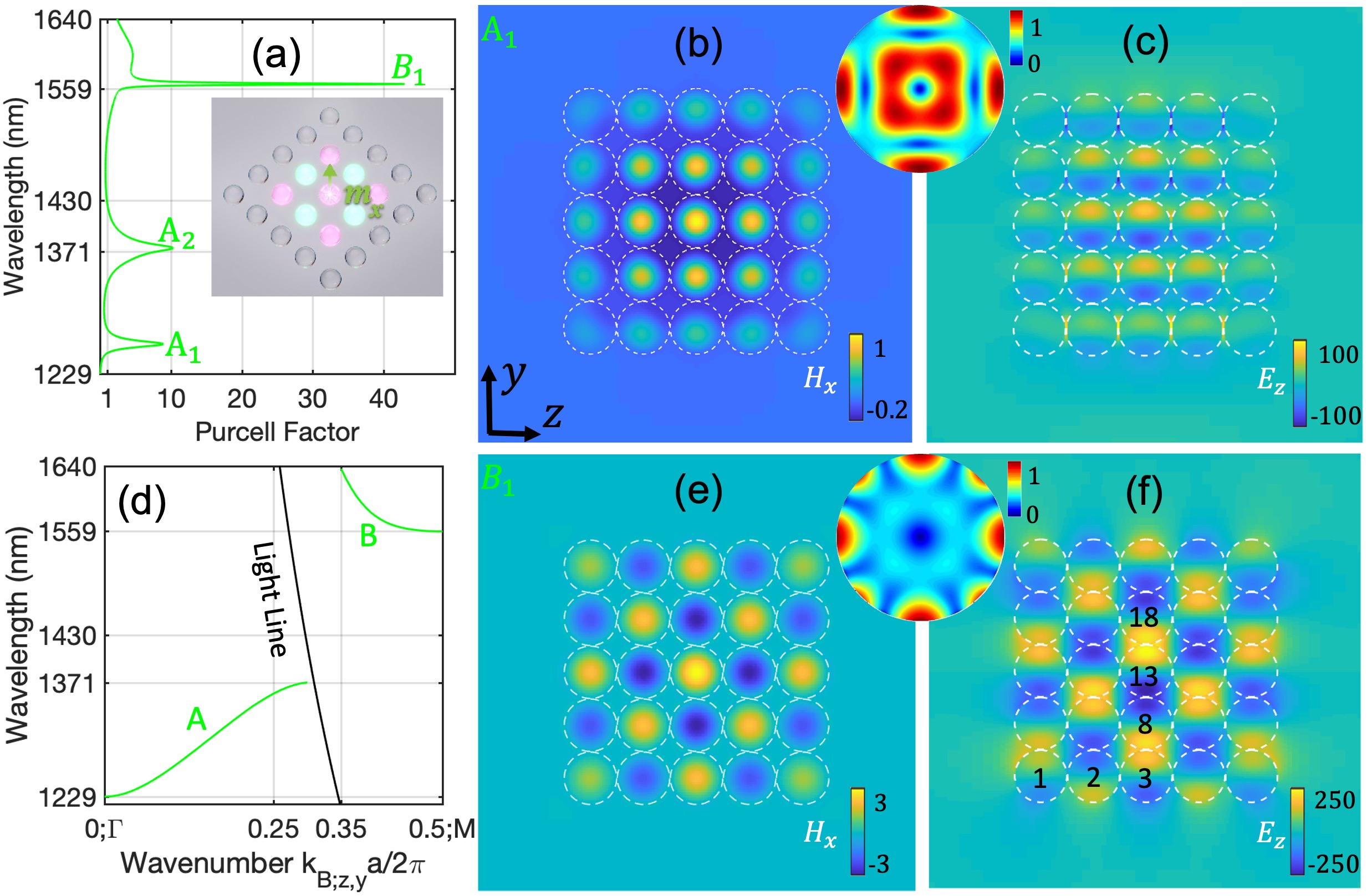}% Here is how to import EPS art
\caption{\label{F3} Mie-tronics origin of supermodes in finite arrays and their correspondence to Bloch bands of the infinite photonic crystal.  
(a) Purcell-factor spectrum for the configuration in the inset, revealing two well-separated families of collective resonances (anti-bonding and bonding) with pronounced supermodes labeled $A_{1,2}$ and $B_1$.  
(b),(c) Magnetic ($H_x$) and electric ($E_z$) field distributions of the anti-bonding supermode $A_1$; the inset shows the corresponding far-field radiation pattern.  
(d) Bloch bands of the corresponding infinite photonic crystal, showing the bands labeled $A$ and $B$ associated with the anti-bonding and bonding supermodes in (a). The vertical axis is plotted in wavelength; accordingly, the bands $A$ and $B$ lie above and below the light line when interpreted in the conventional frequency-based representation.
(e),(f) Same as (b),(c), but for the bonding supermode $B_1$. The strong electric-field hotspot in the gap between the 13th and 18th spheres indicates efficient coupling of this collective MD mode to a $z$-oriented ED source.}
\end{figure*}

Figure~\ref{F3}(a) presents the Purcell-factor spectrum for the configuration shown in the inset, where a MD source $m_x$ couples to a $5\times5$ array of spheres. Two distinct families of collective resonances appear on either side of the peak wavelength (1542 nm) of the intrinsic MD mode shown in Fig.~\ref{F2}. Each family contains well-defined supermodes labeled $A_1$, $A_2$, and $B_1$. The $A_1$ and $B_1$ supermodes correspond to the $\Gamma$- and $M$-point edges of the anti-bonding ($A$) and bonding ($B$) Bloch bands in Fig.~\ref{F3}(d), which emerge in the limit of an infinite array. In the photonic-crystal framework, $A_1$ and $B_1$ correspond to a guided resonance and a guided mode, respectively. To obtain the dispersions in Fig.~\ref{F3}(d), the dipole $m_x$ is positioned at the sphere center, and Bloch boundary conditions are applied along the $y$ and $z$ directions.

Because the multipole modes form an orthogonal basis, the source initially excites only the MD resonance. For this mode, the internal field at the sphere center is determined by the induced MD moment $\eta_x = \sqrt{\tfrac{3}{8\pi}} \lambda H_x$, where $H_x$ is the local magnetic field~\cite{Hoang2022high}. The field in each unit cell, however, generally contains contributions from multiple multipoles due to inter-particle coupling. Linton \textit{et al.} combined multipole expansions with Bloch boundary conditions to describe such multipolar interactions in periodic lattices~\cite{Linton2013electromagnetic}. This semianalytical approach explains the quantitative agreement between photonic-crystal band-edge simulations and Mie-tronics supermodes in sufficiently large arrays~\cite{Hoang2020collective}. This agreement holds even though Mie-tronics is formulated in three-dimensional, fully open systems with an infinite number of radiation channels.

The anti-bonding and bonding characters of these supermodes are evident from their near- and far-field profiles. Figures~\ref{F3}(b,c) show the $A_1$ (anti-bonding) supermode, while Figs.~\ref{F3}(e,f) show the $B_1$ (bonding) supermode. The in-plane electric-field ($E_z$) distributions reveal that $A_1$ exhibits nodes between adjacent spheres, consistent with in-phase oscillation of the fields [Fig.~\ref{F3}(b)]. In contrast, $B_1$ displays pronounced hotspots within the interparticle gaps (e.g., between the 13th and 18th spheres in Fig.~\ref{F3}(f)), corresponding to out-of-phase oscillations in neighboring spheres [Fig.~\ref{F3}(e)]. These enhanced local fields enable strong coupling to emitters or detectors oriented along the $z$ direction. Such collective modes therefore provide a platform for mediating optical interactions between integrated transmitters and receivers and may enable photonic networks for emulating interacting physical systems~\cite{Nielsen2010quantum,Berloff2017realizing,Sergeeva2025coherent,Deng2025enhancement}.

A further distinction arises from the magnetic-field distributions $H_x$ in Figs.~\ref{F3}(b,e), which are proportional to the MD moments induced in the spheres~\cite{Hoang2022high}. For $A_1$, the magnetic moments are aligned, whereas in $B_1$ they are anti-aligned. As a result, $B_1$ extends its fields further into the surrounding environment, while $A_1$ confines its fields more tightly within the unit cells. This difference in field localization explains the greater robustness of the anti-bonding supermode to perturbations introduced by a quartz substrate, as discussed in Sec.~\ref{SecIII}.

The far-field radiation patterns [insets of Figs.~\ref{F3}(b,e)] further clarify the nature of the two supermodes. The bonding supermode $B_1$ radiates mainly along directions parallel to the metasurface, consistent with its guided-mode character. In contrast, the anti-bonding supermode $A_1$, as a guided resonance, radiates strongly into oblique directions away from the surface. Similar to the hole arrays discussed in the accompanying manuscript~\cite{Hoang2025unconventional}, the $A_1$ supermode produces a vortex beam with suppressed radiation along the surface normal, confirming its collective MD origin. From an applications perspective, the in-plane confinement of $B_1$ makes it especially appealing for on-chip photonic devices that demand minimal radiation losses away from the plane.

Interestingly, neither $A_1$ nor $B_1$ radiates along the surface normal (the $x$ axis), preventing excitation by normally incident plane waves. This symmetry protection follows directly from their near-field profiles. A normally incident $z$-polarized plane wave excites only supermodes with $E_z$ even in $z$, whereas both $A_1$ and $B_1$ exhibit odd symmetry. Similarly, a $y$-polarized plane wave couples only to supermodes with $E_y$ even in $y$, but here $E_y$ is odd, again forbidding coupling. These symmetry constraints can equivalently be interpreted in terms of the parity of the Bloch eigenfields within the unit cell. Such symmetry considerations provide useful guidelines for tailoring diffractive bands through controlled symmetry breaking, as demonstrated in Sec.~\ref{SecIV}. Nevertheless, the collective nature of supermodes allows normal-incidence excitation in sufficiently large arrays. In Fig.~\ref{F3}(a), only three discrete supermodes appear for the $5\times5$ array, but their number increases rapidly with array size. Due to retardation, some of these additional supermodes exhibit $E_z$ distributions along the $z$ axis, enabling coupling to normally incident plane waves~\cite{Hoang2025collective}.

The discrete peaks in Fig.~\ref{F3}(a) arise from distributed scattering with radiative boundary conditions across the finite array. They correspond to high-$Q$ collective supermodes that are not captured by conventional Bloch-wave analysis. In this framework, Bloch waves propagate without scattering in an infinite lattice and form continuous bonding and anti-bonding bands [Fig.~\ref{F3}(d)]. In finite arrays, edge reflections can confine these waves and produce Fabry–P\'erot-type modes. However, the Mie-tronics supermodes originate from distributed multiple scattering rather than edge reflections. The scattered fields emerge predominantly from the interior of the array, not from the edges, as expected for Fabry–Pérot feedback~\cite{Hoang2020collective}. Distinguishing these mechanisms is nontrivial and underpins unconventional light localization with Mie-tronics~\cite{Hoang2025unconventional}.
\\
\subsection{Multipolar analysis of Mie-tronics supermodes}\label{Sec2E}
\begin{figure*}[hbtp]
\includegraphics[width = 16 cm]{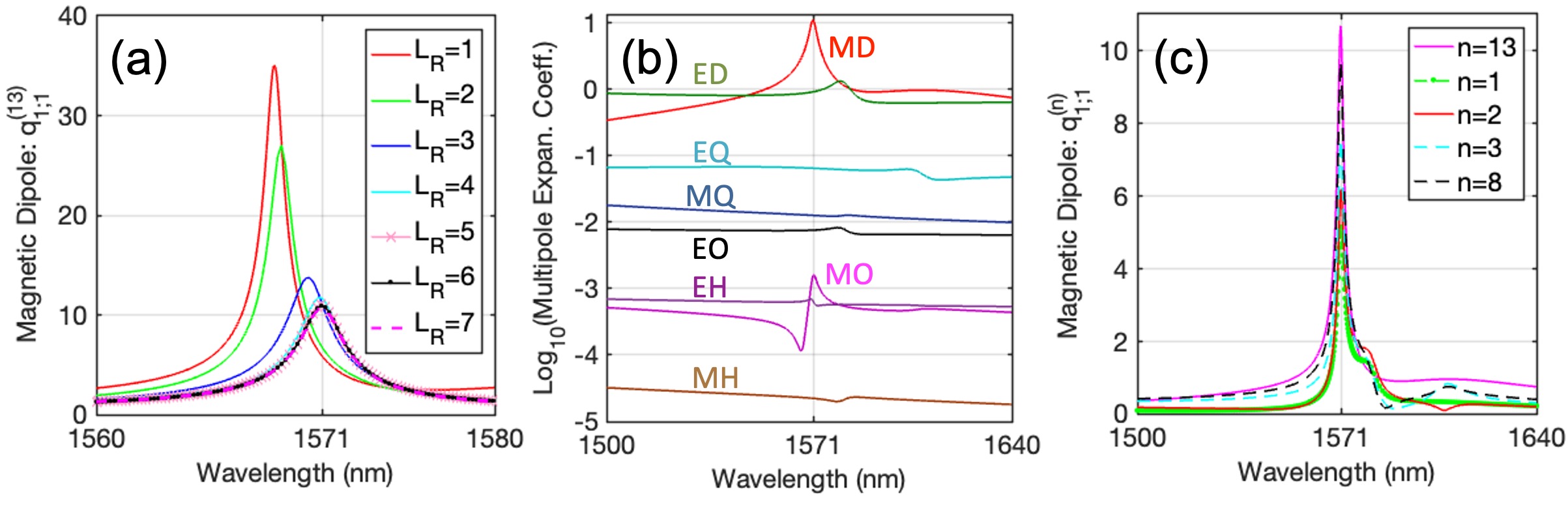}
\caption{\label{F4}
Multipolar analysis of the bonding supermode $B_1$ excited by a $z$-oriented ED source placed in the gap between the 13th and 18th spheres (see Fig.~\ref{F3}(f)). (a) Convergence of the MD coefficient with increasing truncation order $L_R$ of the multipole expansion, showing that higher-order multipoles are required for convergence. (b) Multipole coefficients (electric and magnetic dipole, quadrupole, octupole, and hexapole) of the scattering field from the 13th sphere, confirming the contribution of higher-order multipoles beyond the dominant MD channel (note the logarithmic scale on the $y$ axis). (c) MD coefficients for the 13th, 1st, 2nd, 3rd, and 8th spheres, demonstrating that the resonance occurs collectively across the array at a single peak wavelength.}
\end{figure*}
Recent work has shown that higher-order multipoles are essential for modeling the nonlocal response of metasurfaces, even when their unit cells are deeply subwavelength~\cite{Achouri2023spatial}. In this limit, such structures can be treated as effective homogeneous sheets described by homogenized material models. Here we extend this multipolar framework to finite arrays whose unit cells remain subwavelength but are no longer deeply subwavelength. In this regime, collective resonances emerge from coherent multipolar scattering distributed across the lattice. Such resonances cannot be captured by local effective parameters. Figure~\ref{F4} presents a multipole analysis of the bonding supermode $B_1$ [Fig.~\ref{F3}(f)].

Before quantitatively analyzing the role of higher-order multipoles, we briefly outline the formulation describing light scattering by a group of spheres. Collective resonances are excited by a $z$-oriented ED source located at the origin $O$, with dipole moment $\mu_{1;0}$. The scattered field from the $n$-th sphere at position $O_n$ is expressed using multipole expansion coefficients $p^{(n)}_{l^\prime m^\prime}$ and $q^{(n)}_{l^\prime m^\prime}$. The multipolar responses of the spheres are coupled through the following equations~\cite{Hoang2017fano}:
\begin{widetext}
\begin{align}
    &p^{(n)}_{l^\prime m^\prime}=a_{l^\prime}^{(n)}\left(A_{l^\prime m^\prime}^{1;0}(\overrightarrow{OO_n})\mu_{1;0}+\sum_{u\neq n}\sum_{l=1}^{L_u}\sum_{m=-l}^l\left[A_{l^\prime m^\prime}^{lm}(\overrightarrow{O_uO_n})p_{lm}^{(u)}+iB_{l^\prime m^\prime}^{lm}(\overrightarrow{O_uO_n})q_{lm}^{(u)}\right]\right)\label{E2}\\
    &q^{(n)}_{l^\prime m^\prime}=b_{l^\prime}^{(n)}\left(-iB_{l^\prime m^\prime}^{1;0}(\overrightarrow{OO_n})\mu_{1;0}+\sum_{u\neq n}\sum_{l=1}^{L_u}\sum_{m=-l}^l\left[A_{l^\prime m^\prime}^{lm}(\overrightarrow{O_uO_n})q_{lm}^{(u)}-iB_{l^\prime m^\prime}^{lm}(\overrightarrow{O_uO_n})p_{lm}^{(u)}\right]\right).\label{E3}
\end{align}
\end{widetext}
The coupling coefficients $A_{l^\prime m^\prime}^{lm}$ and $B_{l^\prime m^\prime}^{lm}$ arise from the translational addition theorem, which relates multipole expansions defined in different coordinate systems~\cite{Chew1993efficient}. These coefficients describe wave interference at each scatterer and capture the cooperative nature of light scattering within the array. The formulation therefore explicitly accounts for nonlocal interactions between spheres, including both dipolar and higher-order multipolar contributions.

We now turn to quantitative results for the $5\times5$ array of identical spheres [Fig.~\ref{F4}]. Because all the spheres have the same radius $R$, the truncation order is identical for each sphere ($L_u=L_R$ as defined in Eq.~\eqref{E1}). Figure~\ref{F4}(a) shows the convergence of the MD response as higher-order multipoles are included in the model defined by Eqs.~\eqref{E2}–\eqref{E3}. The MD spectrum associated with the central sphere converges only when the truncation order reaches $L_R=5$. Increasing $L_R$ introduces additional radiation channels that progressively broaden the resonant peak.

We examine the multipolar composition of the resonance by plotting representative coefficients from the 13th sphere in Fig.~\ref{F4}(b): MD, ED, electric quadrupole (EQ), magnetic quadrupole (MQ), electric octupole (EO), magnetic octupole (MO), electric hexadecapole (EH), and magnetic hexadecapole (MH). Each multipole order $l$ contains $2l+1$ modes ($m=-l,\ldots,l$); here one representative coefficient is shown for clarity. MATLAB code for calculating all the multipole coefficients is publicly available~\cite{Zenodo}. Because higher-order multipoles are off-resonant in this wavelength range, their amplitudes remain orders of magnitude smaller than those of the lower-order modes.

The broad spectral overlap of MD and ED modes shown in Fig.~\ref{F2} is also evident in Fig.~\ref{F4}(b). Near 1500~nm, the ED contribution becomes comparable to, and eventually exceeds, the MD as the system approaches the regime of collective ED resonances. In contrast, higher-order multipoles vary smoothly with wavelength and do not exhibit sharp peaks at the MD resonance. They therefore act as continuous radiation channels, increasing the radiative leakage of the MD supermode.

For spherical resonators, supermodes associated with different Mie resonances, such as ED and MD, typically occur in distinct wavelength ranges. Each mode responds differently to geometric modifications depending on its spatial overlap with the perturbation. By judicious symmetry breaking, the resonator geometry can be engineered to control the spectral overlap of Mie modes. This control can lead to merging of collective resonances, resulting in large $Q-$factors and enhanced Purcell effects~\cite{Hoang2025unconventional}.

Because collective resonances arise from distributed scattering across the array, the field enhancement is strongest at the central unit cell [Fig.~\ref{F3}(e)]. Figure~\ref{F4}(c) shows the MD magnitudes for five spheres marked in Fig.~\ref{F3}(f). The amplitude decreases with distance from the center, while the resonance wavelength remains identical for all spheres. This behavior confirms that the supermode originates from phase-coherent multipolar coupling across the array rather than from independent localized resonances of individual scatterers.

These results show that accurate modeling requires higher-order multipoles, even for subwavelength unit cells. As additional multipolar channels open, they increase radiative leakage and set the resonance linewidth of supermodes. In the next section, we examine how symmetry and symmetry breaking influence their formation and $Q-$factors.

\section{Impact of Symmetry Breaking on Mie-tronics Supermodes}\label{SecIII}
Symmetry breaking of unit cells has been extensively explored for engineering diffractive bands. Its impact on collective resonances in finite metasurfaces, however, remains far less understood. Here we show that symmetry breaking not only modifies the coupling between Mie-tronics supermodes and external radiation but can also produce counterintuitive $Q-$factor enhancement—opposite to the trends predicted by infinite-lattice Bloch theory. We also show that the distinct field distributions of Mie modes provide a systematic basis for designing nonlocal metasurfaces with tailored confinement, coupling, and polarization properties.

\subsection{Robustness of anti-bonding supermodes against symmetry breaking}

\begin{figure*}[htbp]
\includegraphics[width=16 cm]{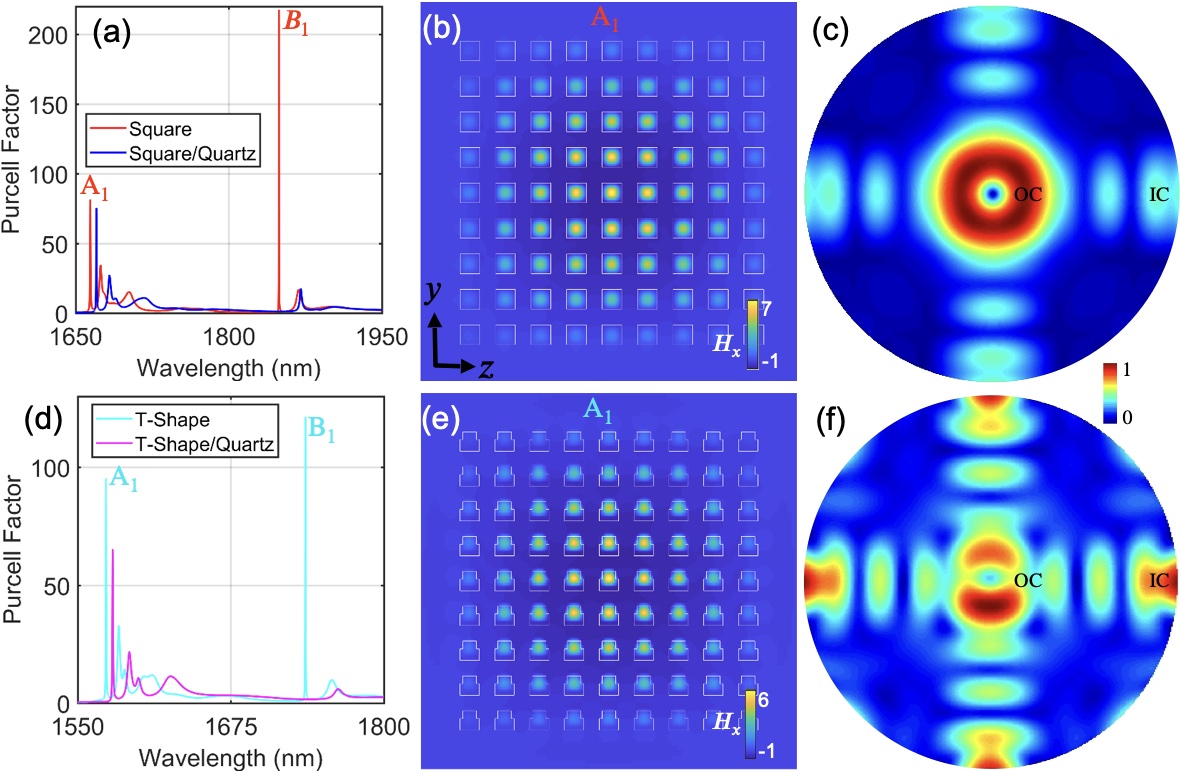}
\caption{\label{F5} Symmetry breaking enhances in-plane multiple scattering, preserves anti-bonding supermodes, and suppresses bonding counterparts in square and T-shaped arrays. 
(a) Purcell-factor spectrum for a magnetic dipole $m_x$ at the center of a $9\times9$ array of square unit cells. The presence of a quartz substrate (refractive index $\simeq 1.45$) suppresses the bonding supermode $B_1$, while the anti-bonding supermode $A_1$ remains nearly unaffected. 
(b),(c) Magnetic near-field distribution and vortex-like far-field pattern of $A_1$. The strong oblique radiation is consistent with its guided-resonance character.   
(d)–(f) Same as (a)–(c) but for the T-shaped array. The far-field pattern highlights enhanced in-plane multiple scattering along in-plane channels (ICs) versus suppressed oblique channels (OCs).}
\end{figure*}

Figure~\ref{F5} summarizes the effects of horizontal and vertical symmetry breaking on Mie-tronics supermodes. Horizontal symmetry breaking is introduced by replacing square unit cells with T-shaped ones, which preserve the lattice vectors but reduce the unit-cell symmetry. Vertical symmetry breaking is implemented by introducing a quartz substrate. For the free-standing square lattice, two well-separated resonant bands appear [Fig.~\ref{F5}(a)], corresponding to the anti-bonding ($A_1$) and bonding ($B_1$) supermodes at the band edges. The magnetic-field distribution of $A_1$ [Fig.~\ref{F5}(b)] confirms its MD character, consistent with the sphere array [Fig.~\ref{F3}]. Introducing the substrate strongly perturbs $B_1$, whose field distribution extends into the surrounding environment and couples efficiently to substrate-supported channels. In contrast, $A_1$ remains largely unaffected because its fields are tightly confined within the unit cells, limiting interaction with the substrate.

The vortex-like far-field pattern of $A_1$ [Fig.~\ref{F5}(c)] indicates that its energy leaks primarily into oblique directions away from the surface, with weaker radiation along the in-plane directions ($y$ and $z$ axes). Compared with the $5\times5$ sphere array [Fig.~\ref{F3}], the $9\times9$ square array supports more radiation channels, reflecting stronger in-plane multiple scattering in larger systems. This enhanced scattering leads to a higher Purcell factor in Fig.~\ref{F5}(a). We note that the anti-bonding supermode $A_1$ corresponds to a guided resonance above the light line, as discussed in Fig.~\ref{F3}. In contrast, the bonding supermode $B_1$ behaves as a guided mode with mainly in-plane radiation. Guided resonances couple efficiently to free-space radiation and give rise to diffraction bands. In the following, we therefore focus on the anti-bonding supermode $A_1$.

Figures~\ref{F5}(d)–(f) present the response of the T-shaped array. Transforming the square unit cell into a T shape preserves the lattice but breaks its in-plane symmetry, producing a blue shift in the Purcell spectrum [Fig.~\ref{F5}(d)]. Both $A_1$ and $B_1$ supermodes persist in the free-standing configuration and exhibit a substrate response similar to that of the square array. The magnetic near-field distribution [Fig.~\ref{F5}(e)] and far-field pattern [Fig.~\ref{F5}(f)] highlight a key effect of in-plane symmetry breaking: radiation is redistributed from oblique channels (OCs) into in-plane channels (ICs). This redistribution increases the photon dwell time within the metasurface, enhancing the $Q-$factor and setting the stage for the results discussed next in Fig.~\ref{F6}.

\subsection{$Q$-enhancement via symmetry breaking}

\begin{figure*}[htbp]
\includegraphics[width=16 cm]{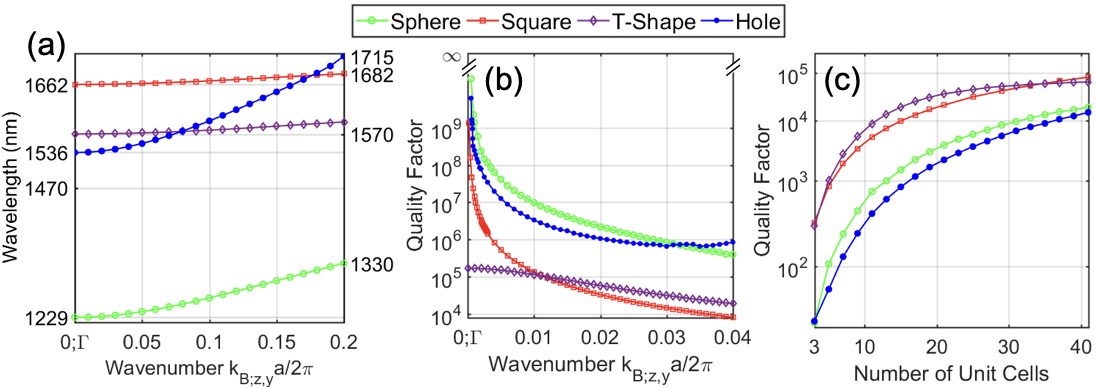}
\caption{\label{F6}
Symmetry breaking lowers the quality factors of Bloch modes in infinite lattices but enhances those of supermodes in finite metasurface arrays.
(a) Anti-bonding Bloch bands above the light line for the four unit-cell types. The band for the sphere unit cell is partially replotted from Fig.~\ref{F3}(d) over a reduced $k$-space range ($k_{B,y}=k_{B,z}\leq 0.2$) for comparison. 
(b) Quality factors of Bloch modes at $\Gamma$: symmetry breaking reduces the divergent value of the square lattice to a finite one in the T-shaped lattice, while sphere and hole lattices show much higher values.
(c) Quality factors of supermodes in finite arrays: symmetry breaking boosts the quality factors of T-shaped arrays beyond that of square arrays for sizes from $5\times5$ to $33\times33$, with both outperforming sphere and hole arrays.}
\end{figure*}

Figure~\ref{F6}(a) shows the anti-bonding Bloch bands above the light line for the four unit-cell geometries. The calculated band-edge wavelengths agree well with the corresponding $A_1$ supermodes (the hole-array case is detailed in the accompanying paper~\cite{Hoang2025unconventional}). Modifying the unit-cell geometry shifts the band edges and alters the dispersion. Notably, the square and T-shaped lattices exhibit narrow spectral widths ($\sim$20~nm), nearly tenfold narrower than that of the hole lattice (179~nm). This result highlights the capability of geometrical design to engineer band dispersion with high precision.

Near the $\Gamma$ point [Fig.~\ref{F6}(b)], the Bloch-mode $Q-$factor diverges for the sphere, square, and hole lattices, reflecting a BIC at the $\Gamma$ point. In contrast, the T-shaped lattice exhibits a finite $Q-$factor, consistent with the conventional picture in which symmetry breaking of unit cell introduces radiative leakage and reduces confinement~\cite{Koshelev2018asymmetric}.

Remarkably, the opposite trend emerges in finite metasurface arrays: symmetry breaking enhances optical confinement by strengthening in-plane multiple scattering. As shown in Fig.~\ref{F6}(c), the T-shaped array achieves the highest $Q-$factor for sizes from $5\times5$ to $33\times33$, reversing the trend predicted by infinite-lattice Bloch analysis. This enhancement results from radiation redistribution, as discussed in the previous subsection [Figs.~\ref{F5}(c) and \ref{F5}(f)]. For very large arrays, however, the proliferation of radiation pathways eventually saturates the $Q-$factor enhancement. This interplay between symmetry breaking and finite-size effects underscores the need to move beyond Bloch-wave descriptions when designing nonlocal metasurfaces for high-$Q$ performance.

Mie-tronics therefore provides practical guidelines for designing finite nonlocal metasurfaces. This design framework is particularly relevant for the miniaturization and integration of light sources and detectors in emerging photonic platforms~\cite{Zhang2026silicon}. In the next section, we show that the modifications of unit-cell geometry and symmetry reshape the diffractive bands of these metasurfaces. 

\section{Diffractive Nonlocal Metasurfaces}\label{SecIV}

The ongoing revolution in artificial intelligence has renewed interest in analog optical computing, driven by its potential for massive parallelism and ultra-low energy consumption. Optical information processing, however, is not new. It has a long history in optical encryption, image processing, and early forms of optical computing~\cite{Ambs2010optical,Chen2014advances}. Around the turn of this century, digital optical computing attracted significant attention. However, progress was ultimately limited by the weak optical nonlinearities of conventional materials, which hindered the realization of practical optical logic gates.

The present resurgence stems from the recognition that artificial intelligence operates through statistical rather than logical principles—tasks inherently suited to analog optical systems~\cite{Mcmahon2023physics,Zhang2025all}. Within this context, diffractive nonlocal metasurfaces have emerged as a promising platform for high-dimensional analog computation. Their collective optical modes, extending across the entire structure, enable both deep miniaturization of on-chip architectures and performance beyond that of conventional optical components~\cite{Guo2020squeeze,Miller2023optics,Wan2024multidimensional,Javidi2025roadmap}.

\subsection{Mie-resonance origin of nonlocal diffractive bands}

\begin{figure}[htbp]
\includegraphics[width=8 cm]{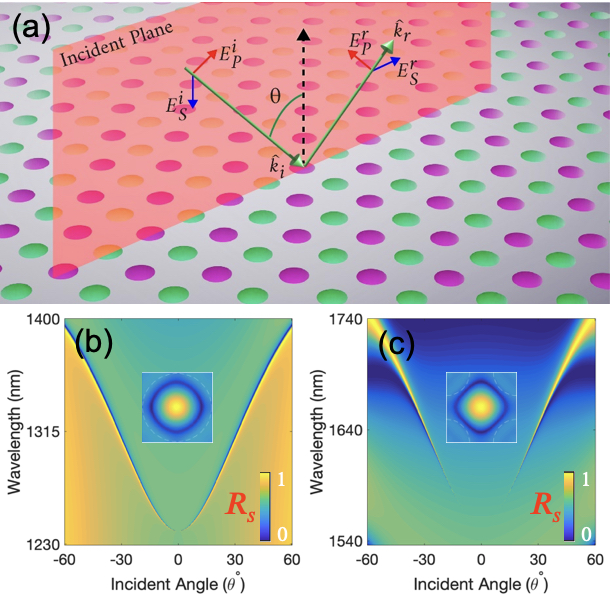}
\caption{\label{F7} Multipolar origin of diffractive nonlocal bands.  
(a) S- or P-polarized plane wave incident on a photonic-crystal slab. 
(b) Reflectance of an S-polarized plane wave from the sphere metasurface, revealing a diffractive band. The inset shows the magnetic field component $|H_x|$ at an incident angle of $5^\circ$. The band vanishes at normal incidence due to symmetry mismatch between the incident field and the eigenmode.
(c) Same as (b) but for the hole metasurface. The inset highlights that the diffractive band originates from the MD resonance.}
\end{figure}

Diffractive nonlocal metasurfaces arise from the interaction between an incident plane wave and an idealized infinite lattice, as schematically illustrated in Fig.~\ref{F7}(a). The plane wave coherently drives all unit cells in phase, exciting their multipolar resonances at specific resonant frequencies (or wavelengths). Each unit cell therefore behaves as a nanoscale resonator supporting discrete multipolar modes, most prominently the ED and MD resonances.

For an isolated sphere, the Cartesian multipoles along the $x$, $y$, and $z$ directions are spectrally degenerate. Coupling within an array lifts this degeneracy, splitting each vectorial multipole into several collective supermodes. For conceptual clarity, we focus on the $x$-oriented MD resonance throughout this work, although dipoles oriented along $y$ or $z$ can also form nonlocal bands under appropriate excitation.

Higher-order multipoles ($l \ge 3$) can likewise produce collective supermodes whose Bloch bands lie entirely above the light line~\cite{Hoang2025collective}. Within diffraction theory, such higher-order bands couple to multiple radiation continua, making them particularly relevant for diffractive nonlocal metasurfaces. Their spectra, however, are densely packed and more difficult to control. Consequently, most analyses—including the present one—focus on the two lowest-order multipoles, MD and ED, whose well-separated resonances allow controlled dispersion engineering through unit-cell geometry.

In coupled arrays, both dipole and quadrupole modes split into bonding and anti-bonding branches. Within diffraction theory, only the anti-bonding supermodes couple efficiently to the radiation continua, giving rise to observable diffractive bands. Because dipole and quadrupole resonances occur at relatively low frequencies, their associated diffraction bands are spectrally well separated and tunable through geometry. Figures~\ref{F7}(b,c) illustrate this effect: reshaping the spherical unit cell into a hole-type geometry transforms the MD-based metasurface from reflection-dominated to transmission-dominated operation. The near-field $H_x$ distributions obtained from RCWA (insets) confirm that these diffractive bands originate from the underlying MD resonance. These MD-based supermodes are commonly referred to as surface lattice resonances~\cite{Garcia2007colloquium,Castellanos2019lattice}.

A key feature of these bands is their disappearance at normal incidence ($\theta = 0^\circ$), commonly referred to as symmetry-protected BICs. Although such BICs are typically associated with singly degenerate diffraction bands, their physical origin is often left unexamined. Mie-tronics provides a clear interpretation by considering how electromagnetic coupling occurs through either electric or magnetic field components~\cite{Awai2007coupling}.

For the $x$-oriented MD mode, magnetic coupling with a P-polarized wave is forbidden because the magnetic field of P polarization lies entirely in-plane ($yz$-plane) and is therefore orthogonal to the MD moment. Electric coupling is likewise prohibited: the in-plane electric field forms a circular displacement current that is antisymmetric with respect to the P-polarization axis ($z$) [Fig.~\ref{F3}(c)], resulting in a zero coupling coefficient. Symmetry breaking can lift this restriction, as demonstrated later for the T-shaped unit cell.

In contrast, an S-polarized wave possesses a magnetic-field component along the normal direction ($x$) for oblique incidence. Consequently, S polarization can excite the MD resonance except at normal incidence ($\theta = 0^\circ$), where the magnetic field has no $x$-component. This symmetry argument explains the polarization selectivity of the MD resonance and clarifies the origin of symmetry-protected BICs in diffractive nonlocal metasurfaces.

While this BIC phenomenon is often described in terms of symmetry selection rules, its physical origin can be understood more fundamentally from the interference of multipolar currents within the unit cell. When the MD eigenmode is excited by a dipole source, multipolar displacement currents are induced in the symmetric unit cell. These currents radiate destructively along the surface normal, forming an eigenmode that cannot couple to a normally incident plane wave~\cite{Hoang2025collective}. This interference mechanism also explains how symmetry breaking enables coupling between P-polarized light and the MD eigenmode, as demonstrated below.

\subsection{Polarization conversion via symmetry breaking}
\begin{figure*}[htbp]
\includegraphics[width=16 cm]{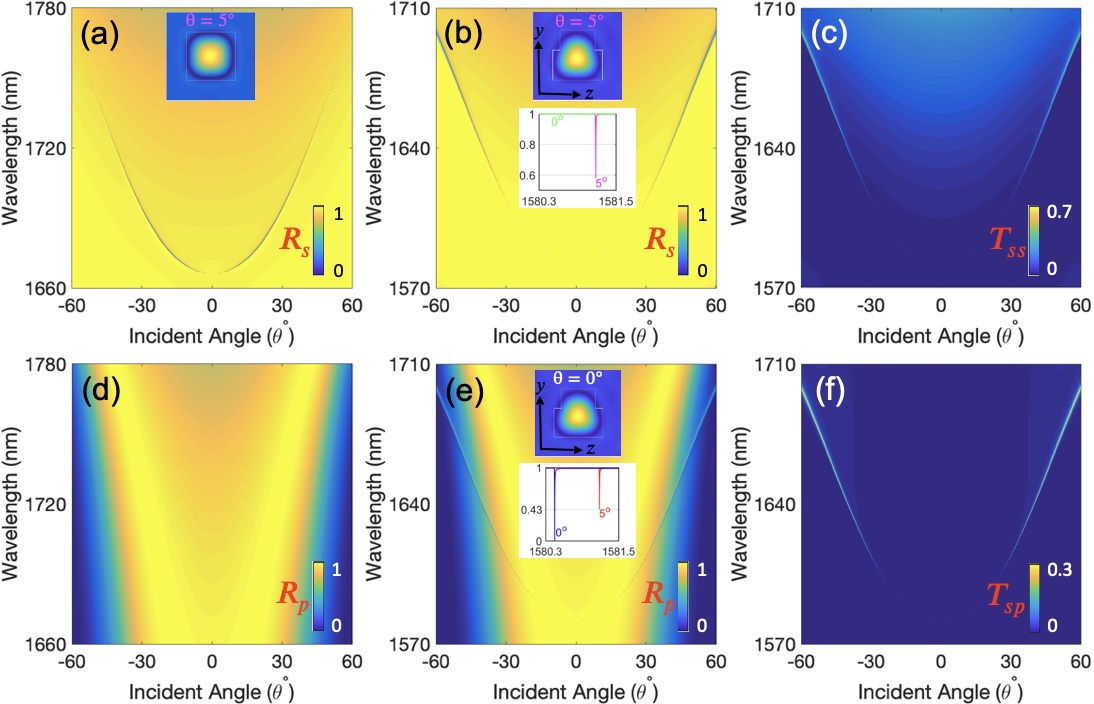}
\caption{\label{F8} 
Symmetry breaking enables polarization conversion in nonlocal metasurfaces.  
(a),(b) Diffractive nonlocal bands in square and T-shaped metasurfaces under S-polarized illumination. Insets show $|H_x|$ at an incident angle of $5^\circ$; both bands vanish at normal incidence, as illustrated in the lower inset of (b).  
(c),(f) Transmittance spectra for S-polarized excitation of the T-shaped metasurface, revealing partial conversion from S to P polarization.  
(d) Reflectance of a P-polarized wave incident on the square metasurface shows no MD-based diffraction band.  
(e) Reflectance of a P-polarized wave incident on the T-shaped metasurface, demonstrating that symmetry breaking enables excitation of the MD-based diffraction band even at normal incidence. The upper inset shows $|H_x|$ at normal incidence ($\theta=0^\circ$). The lower inset shows a sharp drop in reflectance from unity to zero at the eigenmode wavelength. In contrast, for oblique incidence ($\theta=5^\circ$), the reflectance decreases from unity to $\approx$ 0.43.}
\end{figure*}

Breaking in-plane symmetry introduces additional coupling channels that relax polarization selection rules. Figures~\ref{F8}(a)–(b) compare the S-polarized reflectance of square and T-shaped metasurfaces. The near-field distributions [insets of Figs.~\ref{F8}(a)-(b)] confirm that the diffractive bands originate from the MD resonance. The square unit cell behaves similarly to other symmetric unit cells (sphere and hole metasurfaces). Near each diffractive resonance, the S-polarized reflectance ($R_s$) varies rapidly from unity reflection to complete transmission. Furthermore, the diffractive band vanishes at normal incidence due to symmetry mismatch.

Another notable feature is that the S-polarized reflectance of the T-shaped metasurface does not drop from unity to zero across the resonance [inset, Fig.~\ref{F8}(b)], as in symmetric unit cells. This behavior indicates the participation of an additional transmission channel in the diffraction process. Because the lattice period is subwavelength, the only accessible channel is the zeroth-order P-polarized transmission. In other words, the T-shaped metasurface partially converts S-polarized incident light into P-polarized transmission. This polarization conversion is confirmed by the spectra in Figs.~\ref{F8}(c) and \ref{F8}(f). The conversion efficiency reaches up to 30\%. It depends on the degree of symmetry breaking, which governs the electromagnetic coupling between the incident light and the eigenmode. 

Interestingly, despite the symmetry breaking, the diffractive band of the T-shaped array [Fig.~\ref{F8}(b)] still disappears at normal incidence. Because the T-shaped lattice cannot sustain collective resonances with infinite $Q-$factor [Fig.~\ref{F6}], this vanishing response cannot originate from a zero-linewidth mode. Instead, no resonance is excited at $\theta = 0^\circ$. This behavior follows from symmetry considerations. The T-shaped unit cell breaks symmetry only along the $z$ axis and retains mirror symmetry along the $y$ axis. Consequently, the eigenmode’s electric field is antisymmetric along $y$ and is incompatible with the symmetry of the normal incidence. The eigenmode's magnetic field is oriented along $x$ and is orthogonal to that of the incident wave. These symmetry mismatches forbid coupling between the eigenmode and the incident field. 

The disappearance of the diffractive band at $\theta=0^\circ$ is often attributed to the BIC at the $\Gamma$-point in symmetric unit cells [Fig.~\ref{F6}(b)]. However, similar behavior has been observed experimentally in finite nonlocal metasurfaces~\cite{Jie2025near}. This attribution can therefore be misleading. Finite structures cannot support true resonances at the $\Gamma$-point with infinite $Q-$factor. Such states strictly require perfect periodicity. The distinction lies in their physical origins. In finite systems, the disappearance originates from symmetry mismatch due to destructive interference. In contrast, the BIC at the $\Gamma$-point arises from the singularity of wave summation in infinite lattices~\cite{Hoang2025collective}.

Figures~\ref{F8}(d)–(e) further illustrate the effects of symmetry breaking by comparing the P-polarized reflectance of square and T-shaped metasurfaces. P-polarized light cannot excite the MD-based diffraction band in the square array due to the same symmetry constraints discussed earlier. In contrast, the T-shaped array enables coupling between the MD eigenmode and the incident P-polarized wave at all angles. This coupling arises from the overlap of their electric fields along the $z$ axis.  

Under oblique incidence, the previously observed S$\rightarrow$P conversion establishes a coupling between the polarization channels. By electromagnetic reciprocity, this coupling is symmetric. Excitation with P polarization therefore produces an S-polarized response. This is confirmed by the inset of Fig.~\ref{F8}(e), where the reflectance decreases from unity to a nonzero value ($\approx$ 0.43). The behavior is consistent with that observed for S-polarized incidence at $5^\circ$ [inset, Fig.~\ref{F8}(b)]. 

At normal incidence, however, the reflectance drops sharply from unity to zero at the eigenmode wavelength. This indicates the absence of polarization conversion. The MD eigenmode remains symmetry-forbidden from coupling to the S-polarized radiation channel along the surface normal, explaining the disappearance of the diffraction band in Fig.~\ref{F8}(b). Figure~\ref{F8}(e) shows that the diffractive band retains a finite linewidth across all incident angles, reflecting the finite $Q-$factor of the T-shaped metasurface [Fig.~\ref{F6}].

These results show that engineering the symmetry of Mie modes provides a systematic strategy for controlling their coupling to plane waves and for optimizing nonlocal metasurfaces with tunable polarization filtering and conversion.

\section{Discussion and Conclusions}
Metasurfaces have become a cornerstone of modern photonics, offering miniaturized chip-scale platforms for advanced optical functionalities. Yet, pushing their performance beyond current limits requires explicit consideration of nonlocal interactions among unit cells. Conventional metasurface design largely relies on diffractive bands interpreted through Bloch waves, concepts originally developed for infinitely periodic lattices. While Bloch waves underpin much of the success of the electronics industry by providing an intuitive description of scattering-free propagation of electron in periodic lattices, they cannot capture electronic localization in real space. This limitation leads to the introduction of Wannier functions~\cite{Wannier1962dynamics}, which now form the basis of electronic-structure calculations for finite systems~\cite{Marrazzo2024wannier}.

Optical analogues of Wannier functions have been explored for modeling light localization. However, the exponential confinement of electronic Wannier functions does not naturally arise in electromagnetic systems. Instead, as shown in the accompanying work~\cite{Hoang2025unconventional}, multipole functions provide a more rigorous basis for describing light localization and collective resonances in meta-structures. Moreover, the $Q-$factor enhancement arising from symmetry breaking in this work cannot be captured within Bloch-wave descriptions but is naturally explained within the emerging framework of Mie-tronics. In such three-dimensional, fully open systems, the $Q-$factors of supermodes in T-shaped and square arrays exceed those in sphere and hole arrays, in contrast to trends predicted by Bloch-mode analyses. This finding highlights the importance of accounting for emitter–structure interactions within the Mie-tronics framework, particularly since realistic emitters (e.g., quantum dots) radiate into multiple channels. 

The Mie-tronics design framework is built on two key elements: the Mie-mode basis and the multipolar coupling coefficients derived from the addition theorem. The Mie-mode basis describes the electromagnetic structure of individual meta-atoms (unit cells of the metasurface), whereas the coupling coefficients capture both short- and long-range  interactions among them. The importance of long-range interactions has been increasingly recognized in recent years~\cite{Rider2022advances,Li2025realization,Arjas2025topological}. Within this framework, judicious breaking of unit-cell symmetry enables precise control over the coupling between metasurface supermodes and external excitations. Our results show that Mie-tronics provides a unified description of scattering in finite arrays and diffraction in infinite periodic arrays, thereby capturing both localized and extended electromagnetic phenomena.

Advances in high-performance computing now enable direct numerical solutions of Maxwell's equations for large ensembles of Mie scatterers. Building on these computational advances, Mie-tronics offers a systematic framework for analyzing and optimizing both the local behavior of individual meta-atoms and the nonlocal responses of large arrays. Extending this framework to nonlinear, quantum, and time-varying regimes of light–matter interaction will further expand the frontier of nonlocal photonics~\cite{Da2025dynamics}. Such developments may ultimately enable multifunctional integrated photonic circuits that seamlessly integrate computation, control, and communication within a single platform.

\begin{acknowledgments}
This research is supported by the National Research Foundation, Singapore, and A*STAR under its Frontier Competitive Research Programme (NRF-F-CRP-2024-0009), the Australian Research Council (Grant No. DP210101292). D.L. acknowledges support from the Ministry of Education of Singapore under its  SUTD Kickstarter Initiative (Grant No. SKI 20210501). J. Ji acknowledges the funding received from the PhotonDelta National Growth Fund programme.
\end{acknowledgments}

\bibliography{Reference_PRR}% Produces the bibliography via BibTeX.

@article{Hoang2017fano,
  title={Fano resonances from coupled whispering--gallery modes in photonic molecules},
  author={Hoang, Thanh Xuan and Nagelberg, Sara Nicole and Kolle, Mathias and Barbastathis, George},
  journal={Optics Express},
  volume={25},
  number={12},
  pages={13125--13144},
  year={2017},
  publisher={Optica Publishing Group}
}

@article{Hoang2012interpretation,
  title={Interpretation of the scattering mechanism for particles in a focused beam},
  author={Hoang, Thanh Xuan and Chen, Xudong and Sheppard, Colin JR},
  journal={Physical Review A},
  volume={86},
  number={3},
  pages={033817},
  year={2012},
  publisher={APS}
}

@article{Hoang2013rigorous,
  title={Rigorous analytical modeling of high-aperture focusing through a spherical interface},
  author={Hoang, Thanh Xuan and Chen, Xudong and Sheppard, Colin JR},
  journal={JOSA A},
  volume={30},
  number={7},
  pages={1426--1440},
  year={2013},
  publisher={Optica Publishing Group}
}

@article{Hoang2022high,
  title={High-performance dielectric nano-cavities for near-and mid-infrared frequency applications},
  author={Hoang, Thanh Xuan and Chu, Hong-Son and Garc{\'\i}a-Vidal, Francisco J and Png, Ching Eng},
  journal={Journal of Optics},
  volume={24},
  number={9},
  pages={094006},
  year={2022},
  publisher={IOP Publishing}
}

@article{Hoang2020collective,
  title={Collective {M}ie resonances for directional on-chip nanolasers},
  author={Hoang, Thanh Xuan and Ha, Son Tung and Pan, Zhenying and Phua, Wee Kee and Paniagua-Dom{\'\i}nguez, Ram{\'o}n and Png, Ching Eng and Chu, Hong-Son and Kuznetsov, Arseniy I},
  journal={Nano Letters},
  volume={20},
  number={8},
  pages={5655--5661},
  year={2020},
  publisher={ACS Publications}
}

@article{Hoang2024photonic,
  title={Photonic flatband resonances in multiple light scattering},
  author={Hoang, Thanh Xuan and Leykam, Daniel and Kivshar, Yuri},
  journal={Physical Review Letters},
  volume={132},
  number={4},
  pages={043803},
  year={2024},
  publisher={APS}
}

@article{Hoang2025collective,
  title={Collective nature of high-{Q} resonances in finite-size photonic metastructures},
  author={Hoang, Thanh Xuan and Leykam, Daniel and Chu, Hong-Son and Png, Ching Eng and Garc{\i}a-Vidal, Francisco J and Kivshar, Yuri S},
  journal={Physical Review Research},
  year={2025},
  volume={7},
  pages={013316}
}

@article{Liu2019high,
  title={High-{Q} quasibound states in the continuum for nonlinear metasurfaces},
  author={Liu, Zhuojun and Xu, Yi and Lin, Ye and Xiang, Jin and Feng, Tianhua and Cao, Qitao and Li, Juntao and Lan, Sheng and Liu, Jin},
  journal={Physical Review Letters},
  volume={123},
  number={25},
  pages={253901},
  year={2019},
  publisher={APS}
}

@article{Chen2022observation,
  title={Observation of miniaturized bound states in the continuum with ultra-high quality factors},
  author={Chen, Zihao and Yin, Xuefan and Jin, Jicheng and Zheng, Zhao and Zhang, Zixuan and Wang, Feifan and He, Li and Zhen, Bo and Peng, Chao},
  journal={Science Bulletin},
  volume={67},
  number={4},
  pages={359--366},
  year={2022},
  publisher={Elsevier}
}

@article{John1991localization,
  title={Localization of light},
  author={John, Sajeev},
  journal={Physics Today},
  volume={44},
  number={5},
  pages={32--40},
  year={1991},
  publisher={American Institute of Physics}
}

@article{Dorodnyy2023mie,
  title={Mie scattering for photonic devices},
  author={Dorodnyy, Alexander and Smajic, Jasmin and Leuthold, Juerg},
  journal={Laser \& Photonics Reviews},
  volume={17},
  number={9},
  pages={2300055},
  year={2023},
  publisher={Wiley Online Library}
}

@article{Mie1976contributions,
  title={Contributions to the optics of turbid media, particularly of colloidal metal solutions},
  author={Mie, Gustav},
  journal={Annalen der Physik},
  volume={25},
  number={3},
  pages={377--445},
  year={1976}
}

@article{Logan1965survey,
  title={Survey of some early studies of the scattering of plane waves by a sphere},
  author={Logan, Nelson A},
  journal={Proceedings of the IEEE},
  volume={53},
  number={8},
  pages={773--785},
  year={1965},
  publisher={IEEE}
}

@article{Wriedt2012mie,
  title={Mie theory: a review},
  author={Wriedt, Thomas},
  journal={The {M}ie theory: Basics and applications},
  pages={53--71},
  year={2012},
  publisher={Springer}
}

@article{Lagendijk1996resonant,
  title={Resonant multiple scattering of light},
  author={Lagendijk, Ad and Van Tiggelen, Bart A},
  journal={Physics Reports},
  volume={270},
  number={3},
  pages={143--215},
  year={1996},
  publisher={Elsevier}
}

@article{strutt1871lviii,
  title={{LVIII}. {O}n the scattering of light by small particles},
  journal={The London, Edinburgh, and Dublin Philosophical Magazine and Journal of Science},
  volume={41},
  author={Strutt, John William},
  number={275},
  pages={447--454},
  year={1871},
  publisher={Taylor \& Francis}
}

@article{Wannier1962dynamics,
  title={Dynamics of band electrons in electric and magnetic fields},
  author={Wannier, Gregory H},
  journal={Reviews of Modern Physics},
  volume={34},
  number={4},
  pages={645},
  year={1962},
  publisher={APS}
}

@article{Marrazzo2024wannier,
  title={Wannier-function software ecosystem for materials simulations},
  author={Marrazzo, Antimo and Beck, Sophie and Margine, Elena R and Marzari, Nicola and Mostofi, Arash A and Qiao, Junfeng and Souza, Ivo and Tsirkin, Stepan S and Yates, Jonathan R and Pizzi, Giovanni},
  journal={Reviews of Modern Physics},
  volume={96},
  number={4},
  pages={045008},
  year={2024},
  publisher={APS}
}

@article{Rybin2024metaphotonics,
  title={Metaphotonics with subwavelength dielectric resonators},
  author={Rybin, Mikhail V and Kivshar, Yuri},
  journal={npj Nanophotonics},
  volume={1},
  number={1},
  pages={43},
  year={2024},
  publisher={Nature Publishing Group UK London}
}

@article{Chew1993efficient,
  title={Efficient ways to compute the vector addition theorem},
  author={Chew, Weng Cho and Wang, YM},
  journal={Journal of Electromagnetic Waves and Applications},
  volume={7},
  number={5},
  pages={651--665},
  year={1993},
  publisher={Taylor \& Francis}
}

@article{Overvig2020selection,
  title={Selection rules for quasibound states in the continuum},
  author={Overvig, Adam C and Malek, Stephanie C and Carter, Michael J and Shrestha, Sajan and Yu, Nanfang},
  journal={Physical Review B},
  volume={102},
  number={3},
  pages={035434},
  year={2020},
  publisher={APS}
}

@article{Awai2007coupling,
  title={Coupling coefficient of resonators—An intuitive way of its understanding},
  author={Awai, Ikuo and Zhang, Yangjun},
  journal={Electronics and Communications in Japan (Part II: Electronics)},
  volume={90},
  number={9},
  pages={11--18},
  year={2007},
  publisher={Wiley Online Library}
}

@article{Chen2025observation,
  title={Observation of chiral emission enabled by collective guided resonances},
  author={Chen, Ye and Wang, Mingjin and Si, Jiahao and Zhang, Zixuan and Yin, Xuefan and Chen, Jingxuan and Lv, Nianyuan and Tang, Chenyan and Zheng, Wanhua and Kivshar, Yuri and others},
  journal={Nature Nanotechnology},
volume ={20},
  pages={1205},
  year={2025},
  publisher={Nature Publishing Group UK London}
}

@article{Rider2022advances,
  title={Advances and prospects in topological nanoparticle photonics},
  author={Rider, Marie S and Buendia, Alvaro and Abujetas, Diego R and Huidobro, Paloma A and Sanchez-Gil, Jose A and Giannini, Vincenzo},
  journal={ACS photonics},
  volume={9},
  number={5},
  pages={1483--1499},
  year={2022},
  publisher={ACS Publications}
}

@article{Chen2014advances,
  title={Advances in optical security systems},
  author={Chen, Wen and Javidi, Bahram and Chen, Xudong},
  journal={Advances in Optics and Photonics},
  volume={6},
  number={2},
  pages={120--155},
  year={2014},
  publisher={Optical Society of America}
}

@article{Wan2024multidimensional,
  title={Multidimensional encryption by chip-integrated metasurfaces},
  author={Wan, Shuai and Qu, Kening and Shi, Yangyang and Li, Zhe and Wang, Zejing and Dai, Chenjie and Tang, Jiao and Li, Zhongyang},
  journal={ACS nano},
  volume={18},
  number={28},
  pages={18693--18700},
  year={2024},
  publisher={ACS Publications}
}

@article{Javidi2025roadmap,
  title={Roadmap on Optics and Photonics for Security and Encryption},
  author={Javidi, Bahram and Carnicer, Artur and Ahmadi, Kavan and Awatsuji, Yasuhiro and Chen, Wen and Fournel, Thierry and Genevet, Patrice and Guo, Jingying and He, Wenqi and Hebert, Mathieu and others},
  journal={IEEE Access},
  volume={13},
  pages={140087--140117},
  year={2025},
  publisher={Institute of Electrical and Electronics Engineers}
}

@article{Babicheva2024mie,
  title={Mie-resonant metaphotonics},
  author={Babicheva, Viktoriia E and Evlyukhin, Andrey B},
  journal={Advances in Optics and Photonics},
  volume={16},
  number={3},
  pages={539--658},
  year={2024},
  publisher={Optica Publishing Group}
}

@article{Kildishev2025art,
  title={The art of finding the optimal scattering center(s)},
  author={Kildishev, Alexander V and Achouri, Karim and Smirnova, Daria},
  journal={Advanced Optical Materials},
  volume={13},
  number={4},
  pages={2402787},
  year={2025},
  publisher={Wiley Online Library}
}

@article{Krasnok2012all,
  title={All-dielectric optical nanoantennas},
  author={Krasnok, Alexander E and Miroshnichenko, Andrey E and Belov, Pavel A and Kivshar, Yuri S},
  journal={Optics Express},
  volume={20},
  number={18},
  pages={20599--20604},
  year={2012},
  publisher={Optical Society of America}
}

@article{Barati2025toward,
  title={Toward the meta-atom library: experimental validation of machine learning-based {M}ie-tronics},
  author={Barati Sedeh, Hooman and George, Renee C and Lai, Fangxing and Li, Hao and Li, Wenhao and Zheng, Yuruo and Tstekov, Dmitrii and Gao, Jiannan and Moore, Austin and Frantz, Jesse and others},
  journal={Advanced Photonics},
  volume={7},
  number={3},
  pages={036004--036004},
  year={2025},
  publisher={Society of Photo-Optical Instrumentation Engineers}
}

@article{Li2024machine,
  title={Machine learning for engineering meta-atoms with tailored multipolar resonances},
  author={Li, Wenhao and Barati Sedeh, Hooman and Tsvetkov, Dmitrii and Padilla, Willie J and Ren, Simiao and Malof, Jordan and Litchinitser, Natalia M},
  journal={Laser \& Photonics Reviews},
  volume={18},
  number={7},
  pages={2300855},
  year={2024},
  publisher={Wiley Online Library}
}

@article{Guo2018photonic,
  title={Photonic crystal slab {L}aplace operator for image differentiation},
  author={Guo, Cheng and Xiao, Meng and Minkov, Momchil and Shi, Yu and Fan, Shanhui},
  journal={Optica},
  volume={5},
  number={3},
  pages={251--256},
  year={2018},
  publisher={Optical Society of America}
}

@article{Guo2020squeeze,
  title={Squeeze free space with nonlocal flat optics},
  author={Guo, Cheng and Wang, Haiwen and Fan, Shanhui},
  journal={Optica},
  volume={7},
  number={9},
  pages={1133--1138},
  year={2020},
  publisher={Optical Society of America}
}

@article{Kwon2018nonlocal,
  title={Nonlocal metasurfaces for optical signal processing},
  author={Kwon, Hoyeong and Sounas, Dimitrios and Cordaro, Andrea and Polman, Albert and Al{\`u}, Andrea},
  journal={Physical Review Letters},
  volume={121},
  number={17},
  pages={173004},
  year={2018},
  publisher={APS}
}

@article{Komar2021edge,
  title={Edge detection with {M}ie-resonant dielectric metasurfaces},
  author={Komar, Andrei and Aoni, Rifat Ahmmed and Xu, Lei and Rahmani, Mohsen and Miroshnichenko, Andrey E and Neshev, Dragomir N},
  journal={ACS Photonics},
  volume={8},
  number={3},
  pages={864--871},
  year={2021},
  publisher={ACS Publications}
}

@article{Miller2023optics,
  title={Why optics needs thickness},
  author={Miller, David AB},
  journal={Science},
  volume={379},
  number={6627},
  pages={41--45},
  year={2023},
  publisher={American Association for the Advancement of Science}
}

@article{Liu2024edge,
  title={Edge detection imaging by quasi-bound states in the continuum},
  author={Liu, Tingting and Qiu, Jumin and Xu, Lei and Qin, Meibao and Wan, Lipeng and Yu, Tianbao and Liu, Qiegen and Huang, Lujun and Xiao, Shuyuan},
  journal={Nano Letters},
  volume={24},
  number={45},
  pages={14466--14474},
  year={2024},
  publisher={ACS Publications}
}

@article{Brongersma2025second,
  title={The second optical metasurface revolution: moving from science to technology},
  author={Brongersma, Mark L and Pala, Ragip A and Altug, Hatice and Capasso, Federico and Chen, Wei Ting and Majumdar, Arka and Atwater, Harry A},
  journal={Nature Reviews Electrical Engineering},
  volume={2},
  number={2},
  pages={125--143},
  year={2025},
  publisher={Nature Publishing Group UK London}
}

@article{Nguyen2018symmetry,
  title={Symmetry breaking in photonic crystals: on-demand dispersion from flatband to {D}irac cones},
  author={Nguyen, Hai Son and Dubois, Florian and Deschamps, Thierry and Cueff, S{\'e}bastien and Pardon, Antonin and Leclercq, J-L and Seassal, Christian and Letartre, Xavier and Viktorovitch, Pierre},
  journal={Physical review letters},
  volume={120},
  number={6},
  pages={066102},
  year={2018},
  publisher={APS}
}

@article{Koshelev2018asymmetric,
  title={Asymmetric metasurfaces with high-{Q} resonances governed by bound states in the continuum},
  author={Koshelev, Kirill and Lepeshov, Sergey and Liu, Mingkai and Bogdanov, Andrey and Kivshar, Yuri},
  journal={Physical review letters},
  volume={121},
  number={19},
  pages={193903},
  year={2018},
  publisher={APS}
}

@article{Overvig2022diffractive,
  title={Diffractive nonlocal metasurfaces},
  author={Overvig, Adam and Al{\`u}, Andrea},
  journal={Laser \& Photonics Reviews},
  volume={16},
  number={8},
  pages={2100633},
  year={2022},
  publisher={Wiley Online Library}
}

@book{Nielsen2010quantum,
  title={Quantum computation and quantum information},
  author={Nielsen, Michael A and Chuang, Isaac L},
  year={2010},
  publisher={Cambridge university press}
}

@article{Ambs2010optical,
  title={Optical computing: A 60-year adventure},
  author={Ambs, Pierre},
  journal={Advances in Optical Technologies},
  volume={2010},
  number={1},
  pages={372652},
  year={2010},
  publisher={Wiley Online Library}
}

@article{Mcmahon2023physics,
  title={The physics of optical computing},
  author={McMahon, Peter L},
  journal={Nature Reviews Physics},
  volume={5},
  number={12},
  pages={717--734},
  year={2023},
  publisher={Nature Publishing Group UK London}
}

@article{Zhang2025all,
  title={All-optical scalable and programmable VCSEL-based Ising annealer with parallel feedback},
  author={Zhang, Dewen and Yuan, Zifeng and Hoang, Thanh Xuan and Fu, Wujie and Png, Ching Eng and Lim, Soon Thor and Danner, Aaron},
  journal={Optics Express},
  volume={33},
  number={11},
  pages={22119--22131},
  year={2025},
  publisher={Optica Publishing Group}
}

@article{Rayleigh1881x,
  title={X. On the electromagnetic theory of light},
  author={Rayleigh, Lord},
  journal={The London, Edinburgh, and Dublin Philosophical Magazine and Journal of Science},
  volume={12},
  number={73},
  pages={81--101},
  year={1881},
  publisher={Taylor \& Francis}
}

@book{Feynman2015feynman,
  title={The Feynman lectures on physics, Vol. II: The new millennium edition: Mainly electromagnetism and matter},
  author={Feynman, Richard P and Leighton, Robert B and Sands, Matthew},
  volume={2},
  year={2015},
  publisher={Basic Books},
  note={see Chapter~30},
}

@article{Karavaev2025emergence,
  title={Emergence of Diffractive Phenomena in Finite Arrays of Subwavelength Scatterers},
  author={Karavaev, Ilya and Nazarov, Ravshanjon and Li, Yicheng and Bogdanov, Andrey A and Baranov, Denis G},
  journal={Progress In Electromagnetics Research},
  volume={182},
  pages={63--75},
  year={2025}
}

@article{Ho2024finite,
  title={Finite-Area Membrane Metasurfaces for Enhancing Light-Matter Coupling in Monolayer Transition Metal Dichalcogenides},
  author={Ho, Ya-Lun and Fong, Chee Fai and Wu, Yen-Ju and Konishi, Kuniaki and Deng, Chih-Zong and Fu, Jui-Han and Kato, Yuichiro K and Tsukagoshi, Kazuhito and Tung, Vincent and Chen, Chun-Wei},
  journal={ACS nano},
  volume={18},
  number={35},
  pages={24173--24181},
  year={2024},
  publisher={ACS Publications}
}

@article{Shastri2023nonlocal,
  title={Nonlocal flat optics},
  author={Shastri, Kunal and Monticone, Francesco},
  journal={Nature Photonics},
  volume={17},
  number={1},
  pages={36--47},
  year={2023},
  publisher={Nature Publishing Group UK London}
}

@article{Zhang2025vortex,
  title={Vortex lasers through collective boundary scattering},
  author={Zhang, Qiang and Yuan, Xiaocong},
  journal={Nature Nanotechnology},
  volume={20},
  pages={1180},
  year={2025},
  publisher={Nature Publishing Group UK London}
}

@article{Rayleigh1907dynamical,
  title={On the dynamical theory of gratings},
  author={Rayleigh, Lord},
  journal={Proceedings of the Royal Society of London. Series A, Containing Papers of a Mathematical and Physical Character},
  volume={79},
  number={532},
  pages={399--416},
  year={1907},
  publisher={JSTOR}
}

@article{Fano1941theory,
  title={The theory of anomalous diffraction gratings and of quasi-stationary waves on metallic surfaces (Sommerfeld’s waves)},
  author={Fano, Ugo},
  journal={Journal Of The Optical Society Of America},
  volume={31},
  number={3},
  pages={213--222},
  year={1941},
  publisher={Optical Society of America}
}

@article{Liu2012s4,
  title={S4: A free electromagnetic solver for layered periodic structures},
  author={Liu, Victor and Fan, Shanhui},
  journal={Computer Physics Communications},
  volume={183},
  number={10},
  pages={2233--2244},
  year={2012},
  publisher={Elsevier}
}

@article{Hoang2025unconventional,
  title={Unconventional localization of light with {M}ie-tronics},
  author={Hoang, Thanh Xuan and Leykam, Daniel and Nussupbekov, Ayan and Ji, Jie and Rivas, Jaime Gomez and Kivshar, Yuri},
  journal={arXiv preprint arXiv:2507.11995v2},
  year={2025}
}

@article{Cai2020inverse,
  title={Inverse design of metasurfaces with non-local interactions},
  author={Cai, Haogang and Srinivasan, Srilok and Czaplewski, David A and Martinson, Alex BF and Gosztola, David J and Stan, Liliana and Loeffler, Troy and Sankaranarayanan, Subramanian KRS and L{\'o}pez, Daniel},
  journal={npj Computational Materials},
  volume={6},
  number={1},
  pages={116},
  year={2020},
  publisher={Nature Publishing Group UK London}
}

@article{Yao2024nonlocal,
  title={Nonlocal meta-lens with {H}uygens’ bound states in the continuum},
  author={Yao, Jin and Lai, Fangxing and Fan, Yubin and Wang, Yuhan and Huang, Shih-Hsiu and Leng, Borui and Liang, Yao and Lin, Rong and Chen, Shufan and Chen, Mu Ku and others},
  journal={Nature communications},
  volume={15},
  number={1},
  pages={6543},
  year={2024},
  publisher={Nature Publishing Group UK London}
}

@article{Jie2025near,
  title={Near-field Probing of the Local Density of Optical States Enhanced by Bound States in the Continuum in Nonlocal Metasurfaces},
author={Ji, Jie and Sanchez-Gil, Jose and Peeters, Djero and Holman, Wouter and Hoang, Thanh Xuan and van Mechelen, JLM and Gomez-Rivas, Jaime},
  journal={Nature communications},
  volume={16},
  pages={11597},
  year={2025},
  publisher={Nature Publishing Group UK London}
}

@article{You2024resonance,
  title={Resonance wavelength stabilization of quasi-bound states in the continuum constructed by symmetry breaking and area compensation},
  author={You, Shaojun and He, Haoxuan and Zhang, Ying and Duan, Hong and Wang, Lulu and Wang, Yiyuan and Luo, Shengyun and Zhou, Chaobiao},
  journal={Nano Letters},
  volume={24},
  number={48},
  pages={15300--15307},
  year={2024},
  publisher={ACS Publications}
}

@book{Gouesbet2011generalized,
  title={Generalized {L}orenz-{M}ie theories},
  author={Gouesbet, G{\'e}rard and Gr{\'e}han, G{\'e}rard},
  volume={31},
  year={2011},
  publisher={Springer}
}

@article{Won2019into,
  title={Into the ‘{M}ie-tronic’ era},
  author={Won, Rachel},
  journal={Nature Photonics},
  volume={13},
  number={9},
  pages={585--587},
  year={2019},
  publisher={Nature Publishing Group UK London}
}

@book{Bohren2008absorption,
  title={Absorption and scattering of light by small particles},
  author={Bohren, Craig F and Huffman, Donald R},
  year={2008},
  publisher={John Wiley \& Sons}
}

@article{Mao2025lateral,
  title={Lateral, Directional, and Polarized Light Emission from a Silicon Metasurface},
  author={Mao, Yuheng and Zhou, Lidan and Wang, Zhuo and Liu, Shimei and Deng, Fu and Li, Jinxiang and Xiang, Jin and Lan, Sheng},
  journal={Nano Letters},
  volume={25},
  number={36},
  pages={13592--13598},
  year={2025},
  publisher={ACS Publications}
}

@article{Feng2025beyond,
  title={Beyond Fourier harmonics: Anapole-engineered flat-band quasi--bound states in the continuum in dielectric metasurfaces},
  author={Feng, Lutong and Zhang, Xia},
  journal={Physical Review B},
  volume={112},
  number={11},
  pages={115442},
  year={2025},
  publisher={APS}
}

@article{Da2025dynamics,
  title={Dynamics of time-modulated quantum systems via integrated {L}indblad and {M}axwell--{B}loch equations},
  author={da Mota, Achiles F and Sadafi, Mohammad Mojtaba and Chiu, Wei-Chi and Barbiellini, Bernardo and Leuenberger, Michael N and Bansil, Arun and Mosallaei, Hossein},
  journal={APL Quantum},
  volume={2},
  number={3},
  year={2025},
  publisher={AIP Publishing}
}

@article{Ha2024optoelectronic,
  title={Optoelectronic metadevices},
  author={Ha, Son Tung and Li, Qitong and Yang, Joel KW and Demir, Hilmi Volkan and Brongersma, Mark L and Kuznetsov, Arseniy I},
  journal={Science},
  volume={386},
  number={6725},
  pages={eadm7442},
  year={2024},
  publisher={American Association for the Advancement of Science}
}

@article{Berloff2017realizing,
  title={Realizing the classical {XY} Hamiltonian in polariton simulators},
  author={Berloff, Natalia G and Silva, Matteo and Kalinin, Kirill and Askitopoulos, Alexis and T{\"o}pfer, Julian D and Cilibrizzi, Pasquale and Langbein, Wolfgang and Lagoudakis, Pavlos G},
  journal={Nature materials},
  volume={16},
  number={11},
  pages={1120--1126},
  year={2017},
  publisher={Nature Publishing Group UK London}
}

@article{Arjas2025topological,
  title={Topological invariants and topological charges in photonic systems},
  author={Arjas, Kristian and Salerno, Grazia and T{\"o}rm{\"a}, P{\"a}ivi},
  journal={Physical Review B},
  volume={112},
  number={23},
  pages={235428},
  year={2025},
  publisher={APS}
}

@article{Li2025realization,
  title={Realization and manipulation of compact localized states in a two-dimensional photonic crystal with a Lieb lattice},
  author={Li, Haotian and Huang, Renwen and Dong, Renwu and Li, Shiqi and Huang, Hui and Zhang, Xinyang and Chen, Zhuo and Zhan, Peng and Wang, Zhenlin},
  journal={Physical Review Applied},
  volume={23},
  number={5},
  pages={054027},
  year={2025},
  publisher={APS}
}

@article{Barreda2025bound,
  title={Bound states in the continuum in all-dielectric metasurfaces},
  author={Barreda, A and Garc{\'\i}a-Mart{\'\i}n, A and S{\'a}nchez-Gil, JA},
  journal={APL Photonics},
  volume={10},
  number={10},
  year={2025},
  publisher={AIP Publishing}
}

@article{Deng2025enhancement,
  title={Enhancement of single-photon purity and brightness via bound states in the continuum},
  author={Deng, Yating and Fan, Yaqi and Li, Jiahua},
  journal={Physical Review A},
  volume={112},
  number={3},
  pages={033711},
  year={2025},
  publisher={APS}
}

@article{Sergeeva2025coherent,
  title={Coherent Smith--Purcell radiation of a hollow electron beam from a metasurface},
  author={Sergeeva, D Yu and Karlovets, DV and Tishchenko, AA},
  journal={Optics Letters},
  volume={50},
  number={11},
  pages={3724--3727},
  year={2025},
  publisher={Optica Publishing Group}
}

@article{Song2025emergence,
  title={Emergence of cascading flat bands in breathing superlattices},
  author={Song, Moru and Hu, Jinyu and Shi, Lina and Zhang, Yongliang and Chang, Kai},
  journal={Physical Review B},
  volume={112},
  number={8},
  pages={L081401},
  year={2025},
  publisher={APS}
}

@article{kivshar_Mie,
  title={The rise of \uppercase{M}ie-tronics},
  author={Kivshar, Yuri},
  journal={Nano Letters},
  volume={22},
  number={8},
  pages={3513-3515},
  year={2022},
  publisher={APS}
}

@article{Linton2013electromagnetic,
  title={Electromagnetic guided waves on linear arrays of spheres},
  author={Linton, CM and Zalipaev, V and Thompson, Ian},
  journal={Wave Motion},
  volume={50},
  number={1},
  pages={29--40},
  year={2013},
  publisher={Elsevier}
}

@article{Achouri2023spatial,
  title={Spatial symmetries in nonlocal multipolar metasurfaces},
  author={Achouri, Karim and Tiukuvaara, Ville and Martin, Olivier JF},
  journal={Advanced Photonics},
  volume={5},
  number={4},
  pages={046001--046001},
  year={2023},
  publisher={Society of Photo-Optical Instrumentation Engineers}
}

@book{Novotny2012principles,
  title={Principles of nano-optics},
  author={Novotny, Lukas and Hecht, Bert},
  year={2012},
  publisher={Cambridge university press}
}

@misc{Zenodo,
  note = {Hoang, T. X. FDTD and RCWA Simulation Files with MATLAB Codes for Mie-tronics Supermodes and Symmetry Breaking in Nonlocal Metasurfaces. Zenodo, 7 Apr. 2026, https://doi.org/10.5281/zenodo.19454652.}
}

@article{Zhang2026silicon,
  title={In-Silicon Mietronics for Flatband-Enhanced Upconversion and Interfacial Loss Recovery},
  author={Zhang, Yuxiang and Zito, Gianluigi and Gu, Yuyang and Adhikary, Sourav and Tjiptoharsono, Febiana and Wu, Mengfei and Liu, Xiaogang},
  journal={Nano Letters},
  volume={26},
  number={9},
  pages={3291--3297},
  year={2026},
  publisher={ACS Publications}
}

@article{Castellanos2019lattice,
  title={Lattice resonances in dielectric metasurfaces},
  author={Castellanos, Gabriel W and Bai, Ping and G{\'o}mez Rivas, Jaime},
  journal={Journal of Applied Physics},
  volume={125},
  number={21},
  year={2019},
  publisher={AIP Publishing}
}

@article{Garcia2007colloquium,
  title={Colloquium: Light scattering by particle and hole arrays},
  author={Garc{\'\i}a de Abajo, Francisco Javier},
  journal={Reviews of modern physics},
  volume={79},
  number={4},
  pages={1267--1290},
  year={2007},
  publisher={APS}
}

\end{document}